\documentclass[aps,pra,reprint,superscriptaddress,showpacs,showkeys]{revtex4-1}

\usepackage[T1]{fontenc}
\usepackage{graphicx}
\usepackage{amsmath}
\usepackage{textcomp}
\usepackage{xcolor}
\usepackage{soul}

\newif\ifhlchanges
\hlchangestrue % uncomment to highlighting changes
\ifhlchanges
 \colorlet{drkred}{red!80!black} % define dark red color
 \setstcolor{drkred} % set strike though color
 \renewcommand\hl[1]{% redefine highlight command
   \bgroup
   \hskip0pt\color{black}%
   #1%
   \egroup
 }
\else
\colorlet{drkred}{red!80!black} % define dark red color
 \renewcommand\st[1]{\unskip} % overwrite st command
 \renewcommand\hl[1]{% black text in hl command
   \bgroup
   \hskip0pt\color{black}% black
   #1%
   \egroup
 }
\fi
\begin{document}

%\preprint{}

\title{Ultrafast Electron Cooling in an Expanding Ultracold Plasma}


\author{Tobias Kroker}
\email[]{tkroker@physnet.uni-hamburg.de}
\homepage[]{http://www.cui-advanced.uni-hamburg.de/}
%\thanks{}
\affiliation{The Hamburg Centre for Ultrafast Imaging, Luruper Chaussee 149, 22761 Hamburg, Germany}
\affiliation{Center for Optical Quantum Technologies, University of Hamburg, Luruper Chaussee 149, 22761 Hamburg, Germany}

\author{Mario Gro{\ss}mann}
\affiliation{The Hamburg Centre for Ultrafast Imaging, Luruper Chaussee 149, 22761 Hamburg, Germany}
\affiliation{Center for Optical Quantum Technologies, University of Hamburg, Luruper Chaussee 149, 22761 Hamburg, Germany}

\author{Klaus Sengstock}
\affiliation{The Hamburg Centre for Ultrafast Imaging, Luruper Chaussee 149, 22761 Hamburg, Germany}
\affiliation{Center for Optical Quantum Technologies, University of Hamburg, Luruper Chaussee 149, 22761 Hamburg, Germany}

\author{Markus Drescher}
\affiliation{The Hamburg Centre for Ultrafast Imaging, Luruper Chaussee 149, 22761 Hamburg, Germany}
\affiliation{Center for Optical Quantum Technologies, University of Hamburg, Luruper Chaussee 149, 22761 Hamburg, Germany}

\author{Philipp Wessels-Staarmann}
\thanks{Equal contribution}
\affiliation{The Hamburg Centre for Ultrafast Imaging, Luruper Chaussee 149, 22761 Hamburg, Germany}
\affiliation{Center for Optical Quantum Technologies, University of Hamburg, Luruper Chaussee 149, 22761 Hamburg, Germany}

\author{Juliette Simonet}
\thanks{Equal contribution}
\affiliation{The Hamburg Centre for Ultrafast Imaging, Luruper Chaussee 149, 22761 Hamburg, Germany}
\affiliation{Center for Optical Quantum Technologies, University of Hamburg, Luruper Chaussee 149, 22761 Hamburg, Germany}


\date{\today}

\begin{abstract}
 \textbf{\hl{ABSTRACT.} Plasma dynamics critically depends on density and temperature, thus well-controlled experimental realizations are essential benchmarks for theoretical models.
The formation of an ultracold plasma can be triggered by ionizing a tunable number of atoms in a micrometer-sized volume of a \hl{$^{87}$Rb} Bose-Einstein condensate (BEC) by a single femtosecond laser pulse.
The large density combined with the low temperature of the BEC give rise to an initially strongly coupled plasma in a so far unexplored regime bridging ultracold neutral plasma and ionized nanoclusters.
Here, we report on ultrafast cooling of electrons, trapped on orbital trajectories in the long-range Coulomb potential of the dense ionic core, with a cooling rate of 400~\hl{$\mathrm{K}\,\mathrm{ps}^{-1}$}. Furthermore, our experimental setup grants direct access to the electron temperature that relaxes from 5250~K to below 10~K in less than 500~ns.}
\end{abstract}
%\pacs{32.80.Rm, 32.80.Fb, 42.50.Hz, 32.90.+a}
%\keywords{}

\maketitle


\section{Introduction}

Ultrashort laser pulses provide pathways for manipulating and controlling atomic quantum gases on femtosecond time-scales. In particular the strong light-field of a femtosecond laser pulse is able to instantaneously ionize a controlled number of atoms in a Bose-Einstein condensate (BEC). Above a critical number of charged particles, the attractive ionic Coulomb potential is large enough to trap a fraction of the photoelectrons, thus forming an ultracold plasma~\cite{Killian2007}.

\begin{figure}[t]
 \includegraphics[width=1\linewidth]{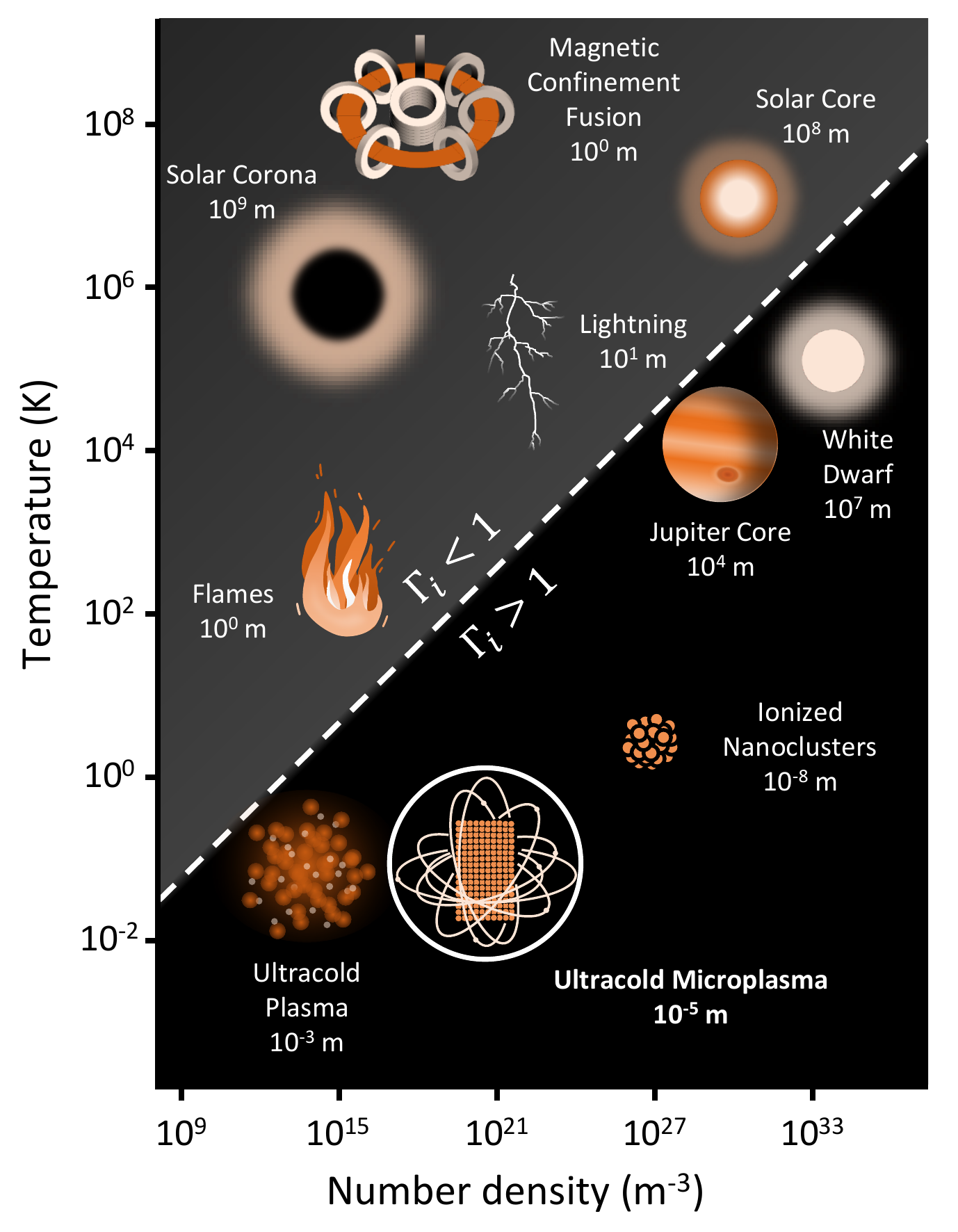}%
 \caption{\label{fig:PlasmaOverview} \textbf{Number density and temperature diagram of plasmas.} Plasmas occurring in nature or prepared in the laboratory span several orders of magnitude in size, temperature and number density. The majority of naturally occurring plasmas are weakly coupled ($\Gamma_\mathrm{i} < 1$), however, the intriguing regime of strongly coupled plasmas ($\Gamma_\mathrm{i} > 1$) is realized in astronomical objects like Jupiter's core or white dwarfs. The dynamics in this challenging region can experimentally be approached by ultracold neutral plasmas, ionized nanoclusters and ultracold microplasmas, where the latter investigated in this work (highlighted by a white circle) bridges the length- and time-scales of the former two.}
\end{figure}

Well-controlled ultracold plasmas in the laboratory provide benchmarks for multi-scale theories and can shed light on extreme conditions present in inertial confinement fusion~\cite{Murillo2004}, the core of Jovian planets and white dwarfs~\cite{Ichimaru1982}. As depicted in Fig.~\ref{fig:PlasmaOverview}, the plasma density inherited from the BEC surpasses the densities achieved so far in supersonic expansion~\cite{Morrison2008} and magneto-optical traps (MOTs)~\cite{Killian1999, Killian2007, Killian2007a, Lyon2017} by orders of magnitude. Strongly-coupled plasmas where the Coulomb energy exceeds the thermal energy are of particular interest because the charge carriers develop spatial correlations~\cite{Viray2020} and self-assembled ordered structures. A recent work~\cite{Langin2019} approaches this regime by laser cooling of the ions.

In small clusters ionized by ultrashort laser pulses, strongly coupled plasmas can be realized as well. In such systems, the interplay between interparticle Coulomb energies and molecular bonds is essential to understand energy transfer between electrons and ions~\cite{Ditmire1998, Saalmann2006, Fennel2010} (see Fig.~\ref{fig:PlasmaOverview}). Recent experiments have studied charged particle dynamics at solid-state densities in finite-size nanoclusters~\cite{Bostedt2010, Gorkhover2012, Kumagai2018} and observed the emergence of low-energy electrons~\cite{Schutte2018}.

\hl{Photoionization of} a BEC with a femtosecond laser pulse \hl{enables access to an unexplored plasma regime} \hl{with} high charge carrier densities above $10^{20}$~m$^{-3}$, cold ion temperatures below 40~mK and hot electron temperatures above 5000~K. Density $\rho$ and temperature $T$ entirely determine not only the coupling parameter but also the dominating length- and time-scales in a plasma. Compared to macroscopic ultracold neutral plasmas (UNP) at MOT densities, these initial parameters allow for creating a micrometer-sized plasma with large charge imbalance and high plasma frequencies where the Coulomb energies initially exceed the ionic thermal energies by three orders of magnitude. Such an initially strongly coupled microplasma with a few hundred to thousands of particles bridges the dynamics and energy transfer studied in photoionized nanoclusters and ultracold neutral plasma. \hl{Moreover, this plasma regime allows neglecting three-body-recombination and interatomic binding energies in the theoretical description, which are relevant in UNP or ionized nanoclusters, respectively. This considerably simplifies theoretical models of the dynamics for benchmark comparisons.}

\hl{Here, we report on the dynamics of ultracold microplasmas triggered in a $^{87}$Rb BEC by a femtosecond laser pulse.} Our experimental setup grants access to the electronic kinetic energy distribution with meV resolution by combining state-of-the-art techniques of ultrashort laser pulses and ultracold atomic gases. So far, the electron temperature of ultracold plasmas has only been inferred indirectly by comparing the fraction of spilled electrons in an extraction field~\cite{Roberts2004}, the free plasma expansion~\cite{Gupta2007} or the three-body recombination rate~\cite{Fletcher2007} to theoretical models. We directly observe an electron cooling from 5250~K to below 10~K in less than 500~ns, which is in excellent agreement with charged particle tracing (CPT) simulations that we have performed in parallel. The small number of particles involved in our microplasma is a key feature that allows for an accurate comparison between experimental results and simulations. \hl{In addition, the} dynamics investigated here reveals striking effects that cannot be captured by an hydrodynamic description such as the ultrafast electron cooling and the increasing electron coupling parameter approaching unity \cite{Killian2007a}.  Such a laboratory experiment, which grants access to \hl{additional}, microscopic observables, allows testing the validity of macroscopic models, thus leading to a better understanding of similar systems in nature.

\section{Results}

\subsection*{Ultracold Microplasma}

\begin{figure}[t]
 \includegraphics[width=1\linewidth]{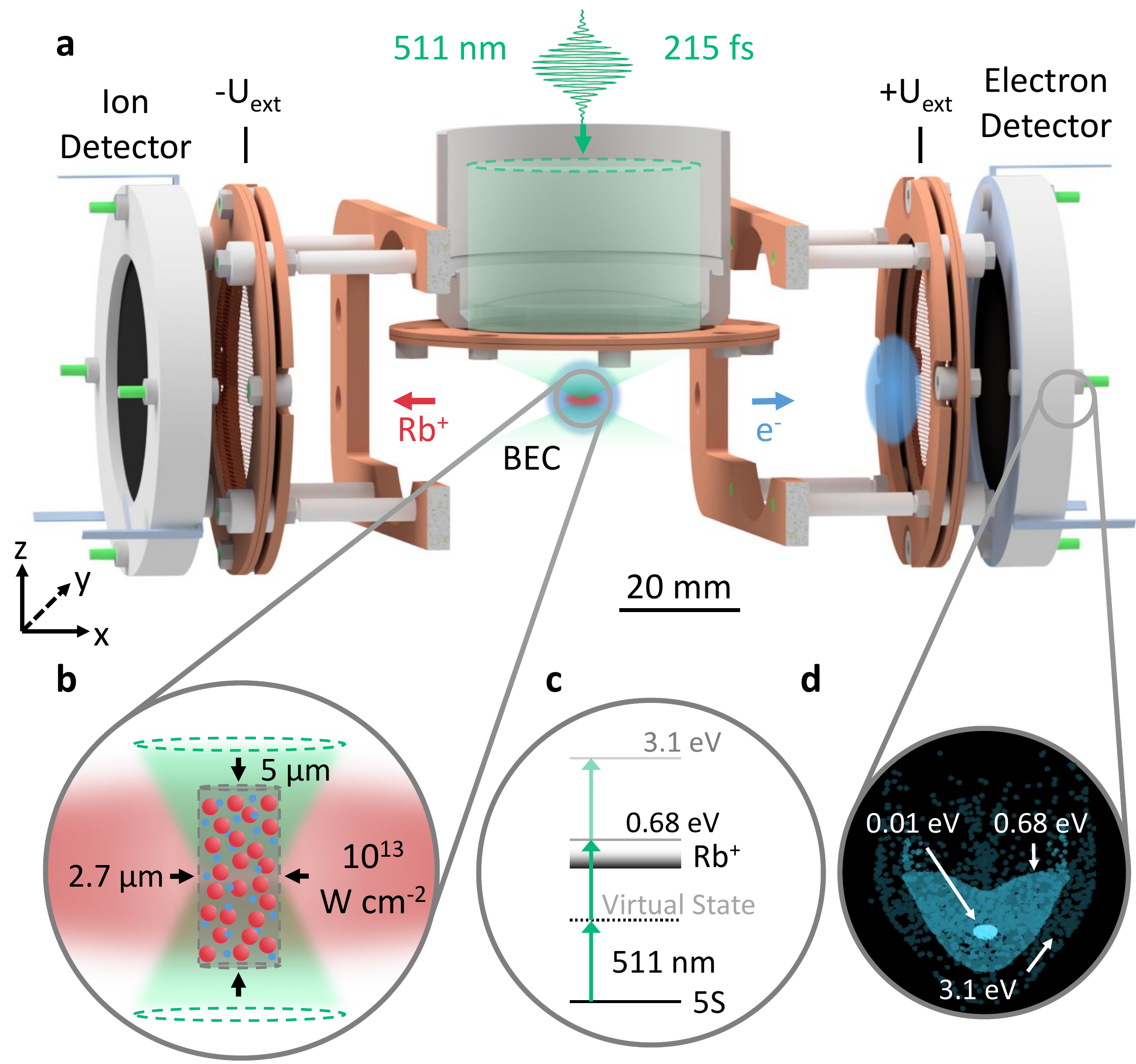}%
 \caption{\label{fig:Setup} \textbf{Detection of ionization fragments after ultrafast ionization of a BEC.} \textbf{a}. A single femtosecond laser pulse locally photoionizes a $^{87}$Rb BEC. The created photoelectrons and ions are separated by an electrical field produced by the two extraction meshes at tunable opposite voltages $\pm U_\mathrm{ext}$. Their kinetic energy distribution is converted into a spatial information during the expansion towards a microchannel plate assembly and a phosphor screen, which is imaged by a highspeed camera. \textbf{b}. At laser peak intensities of $10^{13}$~\hl{$\mathrm{W}\,\text{cm}^{-2}$}, the femtosecond laser pulse ionizes the majority of atoms within a micrometer-sized cylindrical volume creating a charged particle ensemble immersed in the BEC. \textbf{c}. At a central wavelength of 511~nm, $^{87}$Rb is ionized from the $5S$ ground state via non-resonant two- and three-photon processes that correspond to electron excess energies of 0.68~eV and 3.1~eV respectively. \textbf{d}. Simulated electron detector signal after tracing the particle trajectories for three different initial kinetic energies at $\pm U_\mathrm{ext} =$ 300~V. The spatial extent on the detector reflects the kinetic energy distribution of the electrons.}
\end{figure}

We experimentally investigate the dynamics of an ultracold microplasma by combining ultracold quantum gases with the ultrashort timescales of femtosecond laser pulses. As shown in Fig.~\ref{fig:Setup}a, a $^{87}$Rb BEC is locally ionized by a single laser pulse at 511~nm wavelength with a full width at half maximum (FWHM) duration of 215$^{+20} _{-15}$~fs. Whereas earlier photoionization studies in $^{87}$Rb BECs applied nanosecond laser pulses~\cite{Ciampini:BECPhotoionization}, here, the pulse duration is significantly shorter than the timescale for the electron dynamics given by the inverse electron plasma frequency

\begin{equation}
 \omega_\mathrm{p,e}^{-1} = \sqrt{\frac{m_{\mathrm{e}} \epsilon_0}{\rho_{\mathrm{e}} e^2}} = 1.3\text{ ps},
 \label{eq:omega_e}
\end{equation}

\noindent where $m_\mathrm{e}$ is the electron mass and $\rho_\mathrm{e}$ is the initial electronic density. Therefore, the initial plasma dynamics is not perturbed by the laser pulse and the creation of the charged particles can be considered as instantaneous.

A high-resolution objective with a numerical aperture of 0.5 focuses the femtosecond laser pulse down to a waist of $w_0$ $\approx$ 1~\textmu m leading to peak intensities up to $2 \times 10^{13}$~\hl{$\mathrm{W}\,\text{cm}^{-2}$} (see Methods). The number of ionized atoms can be tuned from a few hundred to $N_\mathrm{e,i}$ $\approx$~4000 in a controlled manner via the pulse intensity. At the highest intensity, the ionization probability reaches unity within the center region of a cylindrical volume depicted in Fig.~\ref{fig:Setup}b~\cite{Wessels2018}. The radius of 1.35~\textmu m is determined by the laser focus, while the height of 5 \textmu m is limited by the atomic target, thus providing a locally ionized volume within the atomic cloud (see supplementary \hl{note 1}).

As depicted in Fig.~\ref{fig:Setup}c, the photoionization of $^{87}$Rb at 511~nm can be described as a non-resonant two-photon-process. The excess energy of 0.68~eV is almost entirely transferred to the lighter photoelectrons. Due to the low initial ionic temperature of $T_\mathrm{i}$ $\approx$ 33~mK dominated by the photoionization recoil, we attain a remarkably high initial ionic coupling parameter of $\Gamma_\mathrm{i} = 4800$, which compares the Coulomb energy to the thermal energy per particle:

\begin{equation}
\Gamma_\mathrm{e,i} = \frac{e^2}{4 \pi \epsilon_0 a_\mathrm{e,i} k_\mathrm{B} T_\mathrm{e,i}}.
 \label{eq:Gamma}
\end{equation}

\noindent Here, $T_\mathrm{e,i}$ describes the electron/ion temperature determined by the mean kinetic energy per particle and $a_\mathrm{e,i} = \left( \frac{3}{4 \pi \rho_\mathrm{e,i}} \right)^{1/3}$ denotes the Wigner-Seitz radius at the electron/ion density $\rho_\mathrm{e,i}$. As both the interparticle distance and the kinetic energy of the ions increase during plasma evolution, $\Gamma_\mathrm{i}$ decreases rapidly (see supplementary \hl{note 7}).

As a key feature, this experimental setup grants access to the atomic density via absorption imaging as well as the energy distribution of the photoionization products (Fig.~\ref{fig:Setup}a). A tunable electric field separates electrons and ions and directs them onto opposite imaging microchannel plates (MCP) (see Methods).
Using CPT simulations, the spatial distribution on the detector can be assigned to an electronic kinetic energy distribution in a quantitative manner (see Methods). The simulated detector images for different kinetic energy at $\pm U_\mathrm{ext} = 300$~V are depicted in Fig.~\ref{fig:Setup}d: photoelectrons resulting from two-photon ionization (0.68~eV) or from above-threshold ionization (ATI) processes (3.1~eV)~\cite{Schuricke2011, Hart2016} can be clearly identified.

\subsection*{Measurement Of Electronic Temperature}

\begin{figure*}
 \includegraphics[width=1\linewidth]{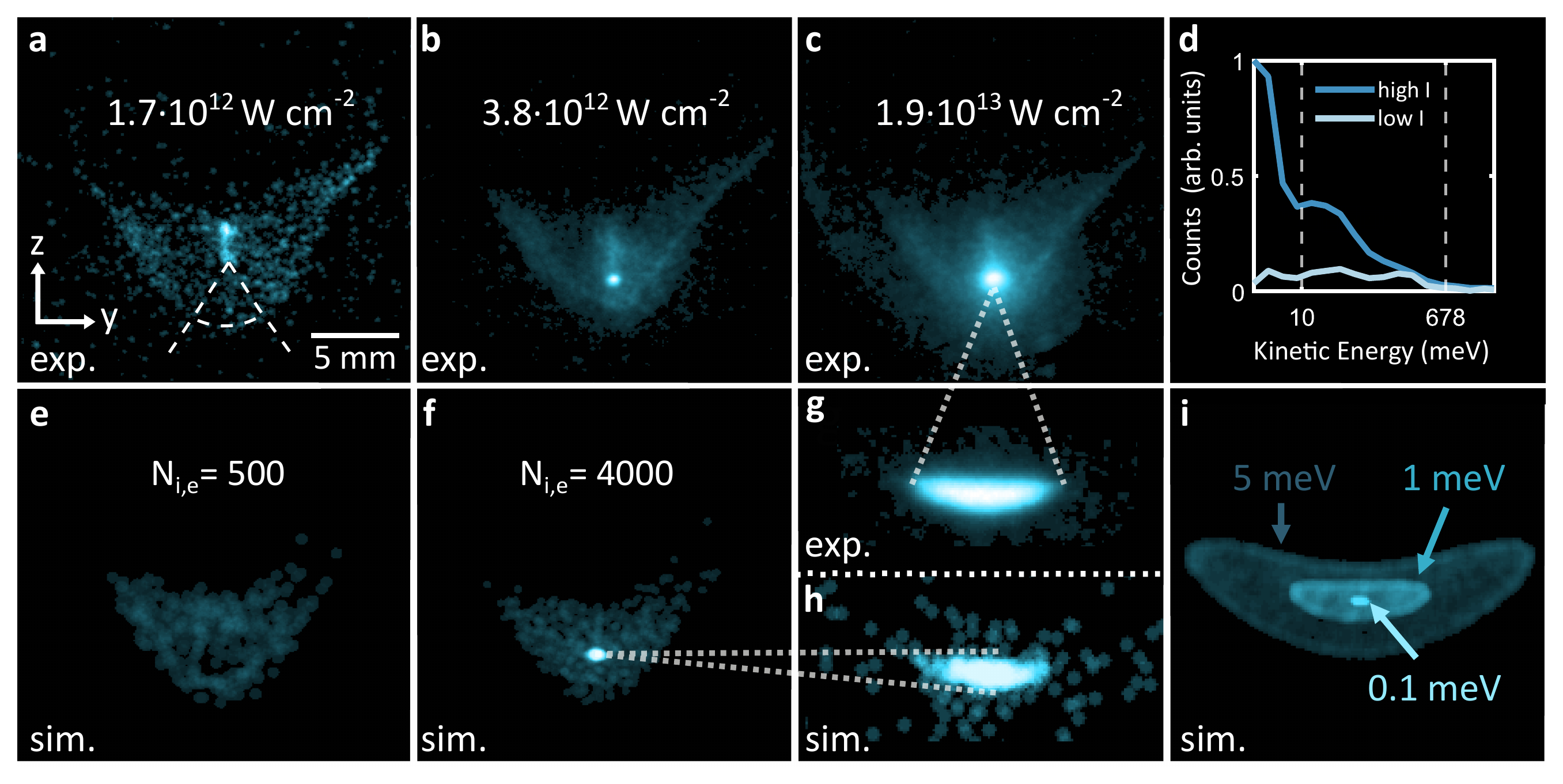}%
 \caption{\label{fig:meas_detector} \textbf{Direct measurements of the electron kinetic energy.} \textbf{a-c}. Measured time-integrated electron distributions at $\pm U_\mathrm{ext} = 300$ V and peak intensity of $I_0 = 0.17, 0.38$ and $1.9 \times 10^{13}$~\hl{$\mathrm{W}\,\text{cm}^{-2}$} (mean detector image over 18, 23 and 22 realizations). A clear signature for the electrons at the two-photon excess energy is obtained (compare Fig.~\ref{fig:Setup}d). As the number of ionized atoms exceeds a critical value, an ultracold plasma is formed signalized here by low kinetic energy electrons (b). At highest intensity, the fraction of slow electrons grows and the signature for three-photon ATI electrons is visible (c). \textbf{d}. Radially averaged spectrum of the recorded electrons at $I_0 = 0.17$ and $0.38 \times 10^{13}$ \hl{$\mathrm{W}\,\text{cm}^{-2}$} within the circular sector depicted in (a). At high intensities, a bright narrow peak appears, corresponding to electrons with kinetic energies below 10 meV. A clear shift of the photoelectrons towards lower energies resulting from the deceleration imposed by the ionic core can be observed. The vertical dashed white lines depict the radii of the electron distributions obtained by trajectory simulations for $E_\mathrm{kin,e}$ = 0.01 and 0.68~eV (see Fig.~\ref{fig:Setup}d). \textbf{e-f}. Plasma simulation results for different electron/ion numbers and an initial kinetic energy of 0.68 eV. For $N_\mathrm{i,e} = 500$ (e) we solely observe the two-photon signature while for $N_\mathrm{i,e} = 4000$ (f) slow electrons are produced as consequence of ultracold plasma dynamics. \textbf{g}. Measured electron distribution for a peak intensity of $I_0$ = $1.2 \times 10^{13}$~\hl{$\mathrm{W}\,\text{cm}^{-2}$} at a lower extraction field of $\pm U_\mathrm{ext} = 5$~V (mean detector image over 14 realizations). \textbf{h}. Electron distribution obtained for $\pm U_\mathrm{ext} = 5$~V extraction field after plasma simulation with $N_\mathrm{i,e}$ = 4000 for an initial electron kinetic energy of $E_\mathrm{kin,e}$ = 0.68~eV. \textbf{i}. Trajectory simulation results for initial kinetic energies of $E_\mathrm{kin,e}$ = 0.1, 1 and 5~meV (same spatial scaling as (g) and (h)).}
\end{figure*}

Figure~\ref{fig:meas_detector}a-c shows the averaged electron signals measured for increasing laser pulse peak intensities. For low intensities (a) the dominant structure on the detector is the spatial distribution of the electrons emerging from the non-resonant two-photon ionization process with a kinetic energy of 0.68~eV, corresponding to an initial electron temperature of $T_\mathrm{e}\approx 5250$~K, in excellent agreement with the trajectory simulation results (Fig.~\ref{fig:Setup}d). For the highest intensity shown (c), a second class of electrons appears stemming from the three-photon ATI (compare to Fig.~\ref{fig:Setup}d).

As a central result, in (b) and (c) an narrow peak appears, corresponding to electrons having a very small kinetic energy. At these intensities, the number of photoionized atoms exceeds the critical number of ions $N^* \approx 960$ required for plasma formation at our excess energies and a fraction of the photoelectrons is trapped and cooled in the resulting space charge potential generated by the unpaired ions. Experimentally, the threshold intensity depends on the density of the atomic target, which gives a clear evidence of a critical number of ions required for plasma formation (see supplementary \hl{note 6}). This rules out low energetic electrons directly created in the strong-field ionization process as reported at high Keldysh parameters~\cite{Blaga2009} or speculated in alkali atoms at high intensities~\cite{Schuricke2011, Pocsai2019}.

Figure~\ref{fig:meas_detector}d shows radially averaged electron distribution in the depicted circular sector in Fig.~\ref{fig:meas_detector}a. The vertical lines mark the limit of the distributions obtained from the trajectory simulations for 0.01 eV and 0.68 eV depicted in Fig.~\ref{fig:Setup}d. At the lowest laser intensity, the kinetic energy distribution is flat up to the energy corresponding to two-photon ionization. At the higher intensity, a large fraction of cold plasma electrons is concentrated below the 10~meV line, which corresponds to a temperature lower than 77 K.

An increasing number of generated ions deepens the space charge potential, which not only enables trapping more electrons in the plasma, but also significantly decelerates the escaping electrons. This can be clearly seen in the averaged spectrum (Fig.~\ref{fig:meas_detector}d) but also as a decrease of the area of the kinetic energy distribution in Fig.~\ref{fig:meas_detector}a-c.

The plasma dynamics is reproduced by CPT plasma simulations including the mutual Coulomb-interactions between all charged particles (see Methods). 
Figure~\ref{fig:meas_detector}e-f shows the simulated results with electron/ion numbers of $N_\mathrm{i,e} =$ 500 (e) and 4000 (f) with an initial electron energy of 0.68~eV, which are in excellent agreement with the measured kinetic energy distributions in Fig.~\ref{fig:meas_detector}a-b. In the simulations slow plasma electrons emerge only above the critical charge carrier density required for plasma formation.

The extraction field sets the expansion time towards the detectors and, thus, the velocity resolution, which can be tuned from 10 meV at $\pm U_\mathrm{ext} = 300$~V to the 1~meV level at $\pm U_\mathrm{ext} = 5$~V (corresponding to static electric fields of 162~\hl{$\mathrm{V}\,\mathrm{m}^{-1}$} and 4.6~\hl{$\mathrm{V}\,\mathrm{m}^{-1}$} in the center, respectively). Figure~\ref{fig:meas_detector}g shows the spatial extent of the low-energy plasma electrons as characteristic elliptical structure for an extraction field of 5 V. The trajectory simulation results for different initial energies are depicted in Fig.~\ref{fig:meas_detector}i. Comparison of Fig.~\ref{fig:meas_detector}g and Fig.~\ref{fig:meas_detector}i yields a measured kinetic energy of the plasma electrons of approximately 1~meV, which corresponds to a final electron temperature below 10~K. 
Figure~\ref{fig:meas_detector}h shows the plasma simulation result for $N_\mathrm{i,e}$ = 4000. Even in this experimentally challenging regime, the measurements and the plasma simulation almost perfectly agree.

\subsection*{Ultrafast Dynamics}

\begin{figure*}
 \includegraphics[width=1\linewidth]{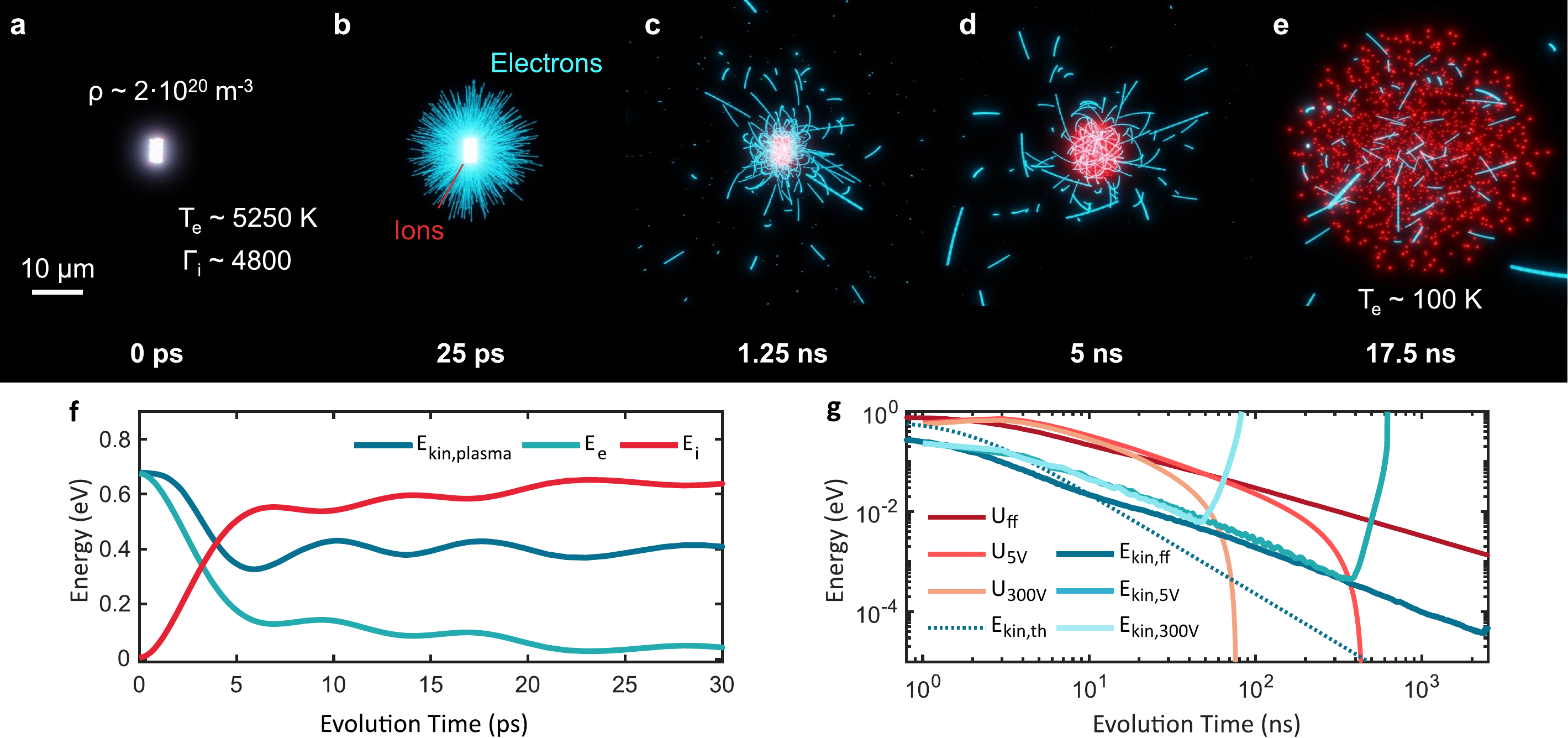}%
 \caption{\label{fig:plasmasketch} \textbf{Ultrafast dynamics of a highly charged ultracold microplasma.} \textbf{a-e}. Snapshots from a CPT simulation tracing the electrons (blue) and ions (red) right after strong-field ionization of an ultracold atomic cloud. An initial homogeneous distribution of ions at a density of $\rho$ = $2 \times 10^{14}$ cm$^{-3}$ and electrons at a temperature of $T_\mathrm{e}$~$\approx$~5250~K is created in a micrometer-sized volume (a). The initial kinetic energy of the ions corresponds to a coupling parameter of $\Gamma_\mathrm{i}$ = 4800 according to Eq.~(\ref{eq:Gamma}). The electrons leave the ionization volume on a picosecond timescale and are subsequently decelerated by the core of remaining ultracold ions (b). While the outer electrons escape, the inner electrons are bound onto orbital trajectories in the ions' attractive Coulomb potential forming an ultracold plasma (c). On a nanosecond timescale the ionic component expands driven by Coulomb-repulsion and further cools the electrons to temperature below 100~K (d-e). \textbf{f}. Picosecond time-evolution of the field-free plasma simulation with the total energy per particle for the electrons (light blue) and the ions (red) and the mean kinetic energy of the electrons captured within the plasma (dark blue). \textbf{g}. Evolution of the kinetic energy of the trapped plasma electrons and the effective depth of the space charge potential on the nano- to microsecond timescale for field-free expansion (dark blue/red), $\pm U_\mathrm{ext} = 5$~V (blue/red) and $\pm U_\mathrm{ext} = 300$~V (light blue/red) as well as the theoretically predicted kinetic energy evolution for adiabatic electron cooling (dotted blue line).}
\end{figure*}

Beyond the excellent agreement with the measured kinetic distributions, these CPT simulations grant access to the dynamics of each particle. They reveal two cooling mechanisms occurring on distinct timescales: an ultrafast cooling during the plasma formation (picosecond timescale) and a subsequent process driven by the Coulomb expansion of the ionic cloud (nanosecond timescale).

Figure~\ref{fig:plasmasketch}a-e shows snapshots of the CPT simulations for an ultracold microplasma consisting of a few thousand charged particles.
While UNP are realized at low excess energies (typically below 0.13~eV)~\cite{Killian1999}, we are able to create an ultracold plasma at high excess energies (0.68 eV), corresponding to an initial electron temperature of $T_\mathrm{e}\approx 5250$~K. Therefore, the majority of photoelectrons leaves the ionization volume within a few picoseconds, while the ions can be regarded as static (Fig.~\ref{fig:plasmasketch}a). 

The decrease of the electronic density in the plasma reduces the amount of shielding between the ions, which thus gain potential energy. In addition, this charge separation process gives rise to a space charge potential that strongly decelerates the escaping electrons (Fig.~\ref{fig:plasmasketch}b). As a result, the kinetic energy of the electronic component is converted to potential energy of the ions. Whereas half of the electrons entirely escape the ionic core (escaping electrons), the other half is trapped within the evoked space charge potential (plasma electrons, see supplementary \hl{note 4} for definition).
Figure~\ref{fig:plasmasketch}f displays the evolution of the mean total electron (light blue line) and ion energy (red line) per particle determined by the sum of kinetic and potential energy of each component. Additionally, the kinetic energy of the plasma electrons (dark blue line) is shown.
Within the electronic expansion, 50\% of the plasma electrons' kinetic energy is transferred onto the potential energy of the ionic component within the first 7~ps.
On this timescale, the trapped electrons are cooled down from 5250~K to about $ 2500$~K, yielding an ultrafast cooling rate for electrons of approximately 400~\hl{$\mathrm{K}\,\mathrm{ps}^{-1}$}. Disorder-induced heating of the electrons is negligible as the associated temperature of $T_\mathrm{DIH} \approx 70~K$ in this density regime is exceeded by orders of magnitude by the initial electron temperature~\cite{Langin2019}.

The large charge imbalance of our plasma strongly influences the many-body dynamics. As depicted in Fig.~\ref{fig:plasmasketch}c, the electrons are trapped in orbital trajectories within the Coulomb potential of a quasi-static ionic core.
This leads to an oscillatory exchange of energy between the captured electrons and the ions (see Fig.~\ref{fig:plasmasketch}f). The period of $2 \pi / \omega_\mathrm{p,e} \approx 8$~ps of these oscillations is consistent with the inverse of the initial electron plasma frequency given in Eq.~\ref{eq:omega_e}. This energy transfer between the individual electrons and the ions is predominantly in phase during the initial electron expansion but it dephases over time leading to a damping behavior (see supplementary \hl{note 4}). In contrast to UNP, here, the ionic plasma period always exceeds the evolution time, preventing ionic thermalization.

On a nanosecond timescale, the potential energy stored in the ions gradually translates into kinetic energy leading to a Coulomb explosion of the plasma (Fig.~\ref{fig:plasmasketch}d,e). 
Whereas UNP typically exhibit hydrodynamic expansion after equilibration due to the electrons thermal pressure~\cite{McQuillen2015}, in this work, the positively charged plasma expansion is dominated by the Coulomb pressure of the charge imbalance, yielding an asymptotic expansion velocity of the root mean square (rms) ion radius of 418~\hl{$\mathrm{m}\,\mathrm{s}^{-1}$} (see supplementary \hl{note 4}). This is in reasonable agreement with the expected hydrodynamic expansion velocity $v_\mathrm{hyd} = \sqrt{k_\mathrm{B} (T_\mathrm{e,0} + T_\mathrm{i,0})/m_\mathrm{i}} = 710$~\hl{$\mathrm{m}\,\mathrm{s}^{-1}$} for the initial electron/ion temperature $T_\mathrm{e/i,0}$~\cite{Killian2007a}.
The ionic expansion leads to a further reduction of the electronic temperature down to $T_\mathrm{e}$ $\approx$ 100~K within tens of nanoseconds. The simulations reveal an increasing electron coupling parameter towards $\Gamma_\mathrm{e} = 0.3$ approaching significant coupling (see supplementary \hl{note 7}).

The evolution of the mean kinetic energy of the plasma electrons as well as the depth of the effective space charge potential during the plasma expansion are shown in Fig.~\ref{fig:plasmasketch}g for CPT simulations for different extraction fields (see supplementary \hl{note 5}). In addition, the expected electron kinetic energy progression given by adiabatic cooling during the plasma expansion $E_\mathrm{kin,e}(t) = E_\mathrm{kin,e}(0) \left(1+ t^2/\tau_\mathrm{exp}^2\right)$ is shown (dotted blue line)~\cite{Killian2007a}. Here, $\tau_\mathrm{exp} = \sqrt{m_\mathrm{i} \sigma^2 / \left[ k_\mathrm{B}\left(T_\mathrm{e,0} + T_\mathrm{i,0}\right) \right]}$ denotes the plasma expansion time and $\sigma$ is given by the initial rms ion radius. The observed electron cooling rates during the plasma expansion largely follow the prediction by the hydrodynamic model. However, the initial ultrafast electron cooling is not captured by this model.

The decrease of the ionic density over time lowers the binding Coulomb potential to the point where its gradient is exceeded by the extraction field~\cite{Twedt2010}. At this point, the plasma electrons are escaping the space charge potential and are drawn to the detector. Hence, the final electron temperatures can be controlled by the extraction field, which determines the plasma lifetime and thus the duration of the electron cooling process.
Without extraction field, electrons can be cooled down to sub meV energies in less than 1~\textmu s.

\subsection*{Plasma Lifetime}

\begin{figure}[t]
 \includegraphics[width=0.92\linewidth]{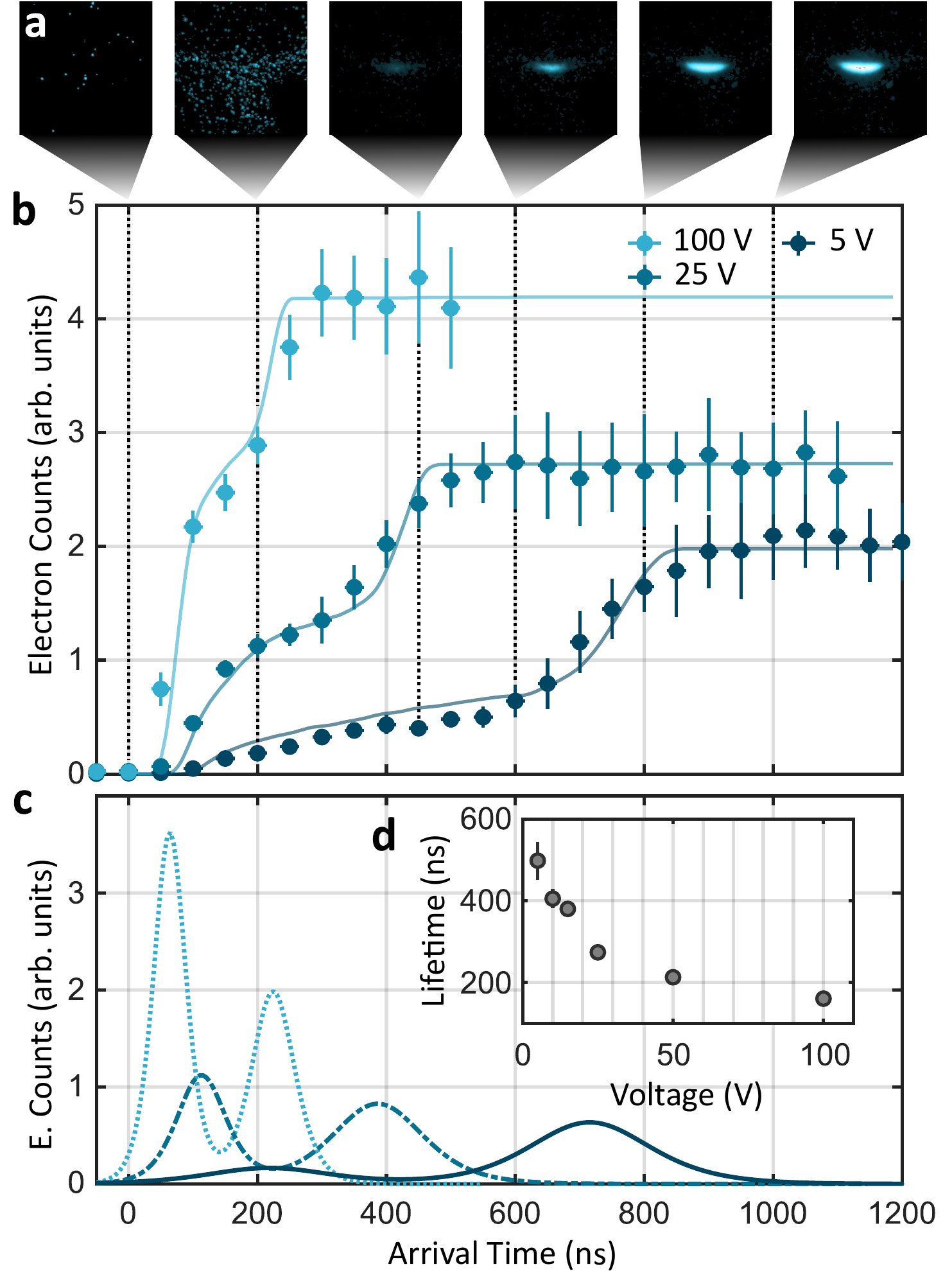}%
 \caption{\label{fig:TofMeasurement} \textbf{Time-resolved photoelectron detection.} \textbf{a}. Cumulated electron signals at an extraction field of $\pm U_\mathrm{ext} = 5$~V for time delays of 0, 200, 450, 600, 800 and 1000~ns after the femtosecond laser pulse (vertical dotted lines in b). The signals are averaged over more than 10 realizations. For a better contrast, the two first images are scaled to their maximum intensity, while the others are normalized to the maximum intensity of the last image. \textbf{b}. Accumulated electron counts measured at peak intensities of $1.2 \times 10^{13}$~\hl{$\mathrm{W}\,\text{cm}^{-2}$} for different extraction fields at $\pm U_\mathrm{ext} = 100$~V (light blue line), $\pm U_\mathrm{ext} = 25$~V (blue line) and $\pm U_\mathrm{ext} = 5$~V (dark blue line) as a function of the maximum arrival time. The counts are spatially integrated over the full detector area. The vertical error bars are given by the standard deviation over all realizations, while the horizontal ones indicate the time uncertainty of the repulsive voltage pulse (see supplementary \hl{note 9}). The solid lines show the vertically scaled results of the corresponding plasma simulations. \textbf{c}. Electron time-of-flight distributions obtained by the derivative of double-sigmoid fits to the data in (b). The two peaks represent escaping and plasma electrons. \textbf{d}. Plasma lifetime for different extraction fields.}
\end{figure}

We determine the lifetime of the microplasma experimentally by implementing a gated detection scheme. In order to separate the slow plasma electrons from the fast escaping electrons, a short repulsive voltage pulse is applied onto the electron extraction electrode after a certain time $t_\mathrm{delay}$ to repel electrons that have not yet passed the electrode (see supplementary \hl{note 9}). The resulting accumulated electron signals for different time delays are depicted in Fig.~\ref{fig:TofMeasurement}a. The escaping electrons arrive on the detector within the first 200~ns and display a homogeneous signal as expected for photoelectrons at 0.68~eV kinetic energies at these low extraction fields. Up to 600~ns, only a small fraction of cold plasma electrons is detected leaking out of the plasma during expansion. In the last 400~ns the fraction of plasma electrons significantly increases as the expanded plasma is torn apart by the extraction field.

For a quantitative analysis, Fig.~\ref{fig:TofMeasurement}b shows the corresponding spatially-integrated electron counts for different extraction voltages. One can clearly distinguish two plateaus that can be associated to the escaping electrons arriving first and the plasma electrons delayed by several hundreds of nanosecond~\cite{Killian1999}.
The arrival time distributions obtained from plasma simulations show excellent agreement with respect to their temporal profile (see Fig.~\ref{fig:TofMeasurement}b - solid lines, vertically scaled with one free parameter).

Figure~\ref{fig:TofMeasurement}c shows the electron arrival rates obtained from the time derivative of a double-sigmoid fit to the data in Fig.~\ref{fig:TofMeasurement}b (see supplementary \hl{note 9}). 
The plasma lifetime corresponds to the time delay between the arrival of escaping and plasma electrons.
The measured lifetimes are plotted for different extraction fields in Fig.~\ref{fig:TofMeasurement}d. 
For the lowest extraction field we achieve a lifetime of \hl{498(46)}~ns, which is in excellent agreement with the calculated vanishing time of the space charge potential in Fig.~\ref{fig:plasmasketch}g.

\section{Discussion}
\label{sec:discussion}

In summary, we have experimentally realized an ultracold plasma in a charge carrier density regime unexplored so far, which supports an initial ion coupling parameter of $\Gamma_\mathrm{i}$~=~4800. We have directly measured electron cooling from 5250~K to below 10~K within 500~ns, arising from the unique combination of a high charge carrier density and a micrometer-sized ionization volume. 
The CPT simulation results are in excellent agreement with the measurements and reveal an ultrafast energy transfer of 50\% of the initial electron energy onto the ionic component within the first plasma period yielding an ultrafast initial electron cooling rate of $\approx$~400~\hl{$\mathrm{K}\,\mathrm{ps}^{-1}$}. 
The \hl{high} degree of experimental control over the initial state, together with the almost perfect agreement of the CPT simulations, provides a unique model system to investigate the validity of statistical plasma models.

Besides the fundamental interest in the plasma dynamics, the low electron temperatures and enormous cooling rates may be used in plasma-based ultracold electron sources~\cite{Claessens2005} (see supplementary \hl{note 11}). The generated low-emittance electron bunches can be utilized for seeding high-brilliance particle accelerators~\cite{McCulloch2011} and coherent imaging of biological systems~\cite{Speirs2015}. Our system links contemporary source designs utilizing MOTs~\cite{Engelen2013, McCulloch2013} where the electron-cathode interactions are negligible and traditional state-of-the-art electron sources~\cite{Zhang2006, Ehberger2015, Zhang2016} where the emittance is fundamentally limited by such interactions. Our density regime indeed promotes an electron cooling mechanism based on their interaction with the ionic core acting as photoemission cathode.

By exploiting the toolbox of ultrafast dynamics further, our experimental setup allows investigating more advanced dynamical schemes. The impact of the plasma geometry can be studied by taking advantage of the non-linearity of the strong-field ionization process in order to shape the ionization volume beyond Gaussian distributions. The interaction between several microplasmas, launched simultaneously within a BEC, can also be explored. Moreover, pump-probe schemes utilizing a second synchronized terahertz pulse for controlling the plasma evolution can offer direct experimental access to the ultrafast dynamics of the microplasma.

The photoionization of a BEC instead of a magneto-optically trapped gas paves the way towards plasmas supporting strong ion as well as electron coupling, mimicking conditions in gas planets~\cite{Ichimaru1982}, confinement fusion~\cite{Murillo2004} or even more exotic systems like quark-gluon-plasma~\cite{Lyon2017}. 
Indeed, strong coupling can be reached by tuning the wavelength of the femtosecond laser close to the ionization threshold, thus reducing the initial kinetic energy of the electrons by orders of magnitude.
Below the ionization threshold, the large spectral bandwidth of the femtosecond laser pulses shall prevent Rydberg blockade effects and enable an efficient excitation of a dense ultracold Rydberg gas, which can form strongly-coupled plasma by avalanche ionization~\cite{Robinson2000, Morrison2008, Mizoguchi2020}.
Finally, disorder-induced heating as limiting process for Coulomb coupling, can be suppressed by loading the condensate into a 3D optical lattice that establishes an initial spatial correlation~\cite{Gericke2003}.

\footnotesize

\section*{Methods}

\subsection*{$^{87}$Rb Bose-Einstein Condensates}

The $^{87}$Rb atoms, evaporated from dispensers, are captured in a 2D MOT used to efficiently load a 3D MOT. Bright molasses cooling allows reaching sub-Doppler temperatures ($T_\text{D} = 148$~\textmu K). The atomic cloud is then loaded into a hybrid trap~\cite{Lin2009} combining a magnetic quadrupole field and a far-detuned dipole trap at 1064~nm and further cooled by forced radio-frequency evaporation. The dipole trap beam transports the ultracold ensemble into the center of the imaging MCP detectors. After further evaporative cooling in a crossed dipole trap, quantum degeneracy can be reached.

A BEC with $N = 4.2 (3) \times 10^{4}$ atoms is held in the far-detuned optical dipole trap at trap frequencies of $\omega_{x,y} = 2 \pi \times 113(3)$~Hz and $\omega_{\mathrm{z}} = 2 \pi \times 128(1)$~Hz providing a peak atomic density of $\rho = 2 \times 10^{14}$~cm$^{-3}$.

\subsection*{Femtosecond Laser Pulses}

In this work, we use laser pulses with a central wavelength of 511.4~nm and a bandwidth of 1.75~nm (FWHM). The duration of the Gaussian temporal profile is 215$^{+20} _{-15}$~fs (FWHM), measured by autocorrelation. A detailed description of the pulse generation can be found in \cite{Wessels2018}.

The pulses are focused by a high resolution microscope objective with a numerical aperture of 0.5 onto the optical dipole trap. The focal waist is measured with an additional, identical objective to $w_1 = 0.99(3)$~\textmu m and $w_2 = 1.00(5)$~\textmu m. The pulse energies $E_\mathrm{p}$ are inferred from the measured averaged laser power $P$ at a pulse repetition rate of 100~kHz, taking the mirror reflectivities as well as the calibrated transmission of the objective and the shielding mesh into account (see supplementary \hl{note 10}). With the measured quantities, the applied peak intensities $I_0$ are determined by

\begin{equation}
 I_0 = \frac{2 P_0}{\pi w_1 w_2}
 \label{eq:peakIntensity}
\end{equation}

\noindent with the peak power $P_0 = E_\mathrm{p}/\left(\sqrt{2 \pi} \tau\right)$ including the rms pulse duration $\tau = \tau_{\mathrm{FWHM}}/\left(2\sqrt{2 \ln(2)}\right)$. The resulting pulse intensity distributions as well as the atomic density distribution allow determining the initial electron/ion distributions within the ionization volume (see supplementary \hl{note 1}).

\subsection*{Charged Particle Detection}

The experimental setup enables direct detection of electrons and ions on opposite spatially resolving detectors. For this purpose, an extraction field accelerates the charged particles onto the detectors. In this work, we analyze the recorded photoelectron kinetic energy distribution. 

The static extraction field is created by two opposing mesh electrodes at $\pm U_\mathrm{ext}$, which consist of gold plated etched copper meshes with about 70-80\% permeability. After passing the meshes, the electrons are post-accelerated towards two MCPs in Chevron configuration with a channel diameter of $12$~\textmu m and a channel pitch of $15$~\textmu m attached to a P46 (Y$_3$Al$_5$O$_{12}$:Ce) phosphor screen. The emitted light from the phosphor screen is recorded by a highspeed camera, which is operated at a frame rate of 1000~fps.
The detection efficiency for electrons $\eta \approx 0.4$ is given by the product of the transparency of the extraction meshes and the open area ratio as well as the quantum efficiency of the MCP. In order to ensure a constant, optimal quantum efficiency for all extraction voltages, the electrons are post-accelerated onto the same front potential $U_\mathrm{front} = 268$~V of the MCP.

Absolute electron numbers are challenging to obtain. Indeed, when decreasing the extraction field, the absolute detection efficiency decreases as electrons with high kinetic energies cannot be detected efficiently. Moreover, the high flux of plasma electrons on a small part of the MCP area as seen for $\pm U_\mathrm{ext} = 300$~V in Fig.~\ref{fig:meas_detector}c leads to electron depletion in the microchannel material and thus a systematic underestimation of the number of incident electrons in this part of the detector.
Furthermore, detection of low kinetic energy electrons is notoriously challenging and typically limited by residual electric and magnetic fields. Therefore, it requires high experimental control over such stray fields (see supplementary \hl{note 3}).

\subsection*{CPT Trajectory Simulations}

In order to obtain a full simulation of the experiment (including the charged particle detectors), trajectory simulations are performed using the Electrostatics as well as the Particle Tracing Module within the COMSOL Multiphysics\textsuperscript{\textregistered} software \cite{COMSOL}. For this reason, we include the 3D computer-aided design (CAD) geometry of our setup into the simulation. The use of finite element methods (FEM) allows calculating the electric potential landscape produced by the different electrodes (see supplementary \hl{note 2}). In addition, a global magnetic offset field of 370~mG perpendicular to the detection axis and the ionization beam axis is included, which is used in the experiment to center the electron signal onto the detector. The trajectory simulation results in Fig.~\ref{fig:Setup}d are obtained for monoenergetic ensembles of electrons having randomized positions within the ionization volume and velocity directions. Due to the expansion during the time-of-flight, the spatial extent of the electron signal grants access to the underlying velocity distribution.

\subsection*{CPT Plasma Simulations}

The CPT simulations of the plasma dynamics are carried out with the Particle Tracing Module within the COMSOL Multiphysics\textsuperscript{\textregistered} software \cite{COMSOL}. For these simulations $N_\mathrm{i,e}$ electrons and ions are randomly distributed in a cylindrical ionization volume. The particles are created monoenergetically according to the two-photon excess energies whereas the distribution of velocity directions is randomized. For $t > 0$, the 3D differential equation of motion is solved numerically for all particles including Coulomb interaction. Typical calculations for $N_\mathrm{i,e} = 4000$ and few microseconds of time-evolution last between 5-22 days while being paralleled on 35 processing units (2.2~GHz) corresponding to a CPU time of a few hundred days.

The divergence of the interparticle Coulomb potential $U_{\mathrm{C}}(r)$ is circumvented by introducing a bounded interaction potential

\begin{equation}
 \tilde{U}_\mathrm{C}(r) =
 \begin{cases}
  U_\mathrm{C}(r), & r > r_0 \\
  U_\mathrm{C}(r_0), & r \le r_0
 \end{cases}  
 \label{eq:boundedCoulomb}
\end{equation}

\noindent where $r$ denotes the interparticle distance and $r_0 = 20$~nm the cut-off radius, which is chosen well below the mean initial interparticle distance. The simulations furthermore neglect interactions with the remaining neutral atoms and radiative energy losses of the charged particles. 

Beyond macroscopic quantities such as the mean kinetic energies of the electron/ion ensembles, the plasma simulations offer detailed access to the dynamics at a single-particle level (see supplementary \hl{note 4})~\cite{Ayllon2019}.

%\subsection*{Data Availability}
%The data that support the findings of this study are available from the corresponding author upon reasonable request.

%\subsection*{Code Availability}
%Any custom computer code used in this study is available from the corresponding author upon reasonable request.

%\bibliography{Ultracold_Plasma}

\begin{acknowledgments}
 \label{sec:acknowledgments}
The authors would like to thank Thomas C. Killian, Nikolay M. Kabachnik and Andrey K. Kazansky for fruitful discussions. We thank Bernhard Ruff, Jakob Butlewski, Julian Fiedler, Donika Imeri and Jette Heyer  for contributions during an early stage of the experiment. This work is funded by the Cluster of Excellence 'CUI: Advanced Imaging of Matter' of the Deutsche Forschungsgemeinschaft (DFG) - EXC 2056 - project ID 390715994 as well as by the Cluster of Excellence 'The Hamburg Centre for Ultrafast Imaging' of the DFG - EXC 1074 - project ID 194651731.
\end{acknowledgments}

%\section*{Author contribution}
%Setup and operation of the experiment was performed by T.K., M.G., J.S., and P.W-S. Measurements, data analysis and simulations were performed by T.K. and M.G. with support from J.S. and P.W-S. The experiment was coordinated and supervised by J.S., P.W-S., K.S., and M.D. The manuscript was written involving all co-authors.

%\section*{Additional information}

%\subsection*{Competing interests}
%The authors declare no competing interests.

%%%% Supplementary %%%%

\normalsize

\clearpage

\appendix

\onecolumngrid
\begin{center}
\textbf{\large Supplementary Material:\\Ultrafast Electron Cooling in an Expanding Ultracold Plasma}
\end{center}
\twocolumngrid

\setcounter{figure}{0}
\setcounter{equation}{0}
\setcounter{table}{0}

\makeatletter
\renewcommand{\thefigure}{S\@arabic\c@figure}
\renewcommand{\theequation}{S\arabic{equation}}
\renewcommand{\thetable}{S\arabic{table}}
\makeatother

\section*{1. Ionization Volume}

In order to calculate the initial electron/ion distributions

\begin{equation}
 \rho_\mathrm{e/i}(x,y,z) = \rho_\mathrm{a}(x,y,z) \times P(x,y,z),
 \label{eq:ionMap}
\end{equation}

\noindent the full 3D distribution of the atomic density $\rho_\mathrm{a}(x,y,z)$ is modeled and multiplied with the non-linear 3D ionization probability distribution $P(x,y,z)$ given by the intensity distribution of the laser pulse. Here, $z$ denotes the pulse propagation direction. The ionization probabilities are obtained by solving the time-dependent Schr\"odinger equation. We have demonstrated in a previous work that this theoretical description is in perfect agreement with the measured ionization probabilities~\cite{Wessels2018}.

%\begin{figure}[tbh]
% \includegraphics[width=1\linewidth]{figS1.PDF}%
% \caption{\label{fig:ionizationVolume} \textbf{Ionization volume}. \textbf{a}. 2D projection of the simulated 3D electron/ion density distribution $\rho_\mathrm{e/i}(x,y,z)$ after strong-field ionization by a single pulse with $I_0 = 1.9 \times 10^{14}$~W/cm$^2$. The dashed white lines mark the cylindrical volume, which is used for the CPT plasma simulations. \textbf{b}. Ionization probability in $x$-direction (solid blue line), in $z$-direction (\hl{red dotted line}) and normalized atomic density distribution in $x$-direction (dashed yellow line).}
%\end{figure}

\begin{figure}[tbh]
 \includegraphics[width=1\linewidth]{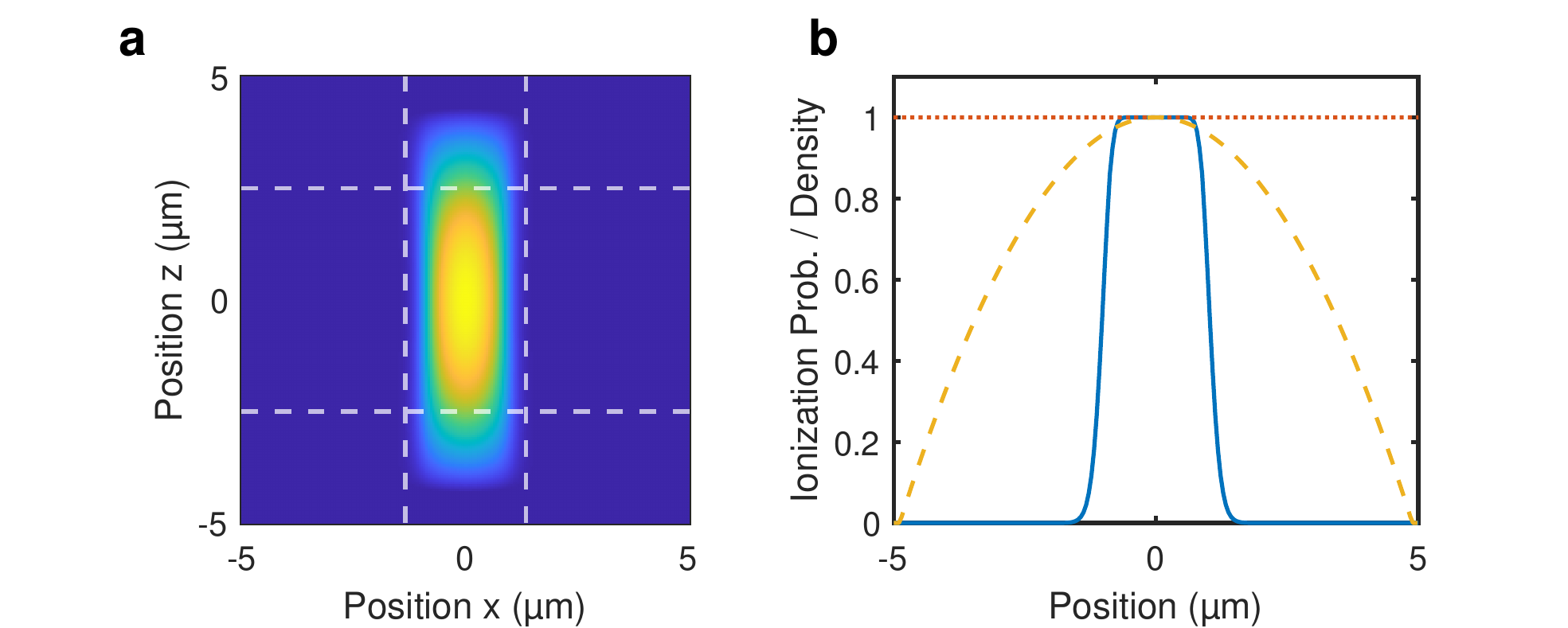}%
 \caption{\label{fig:ionizationVolume} \textbf{Ionization volume}. \textbf{a}. 2D projection of the simulated 3D electron/ion density distribution $\rho_\mathrm{e/i}(x,y,z)$ after strong-field ionization by a single pulse with $I_0 = 1.9 \times 10^{13}$~\hl{$\mathrm{W}\,\text{cm}^{-2}$} The dashed white lines mark the cylindrical volume, which is used for the CPT plasma simulations. \textbf{b}. Ionization probability in $x$-direction (solid blue line), in $z$-direction (red dotted line) and normalized atomic density distribution in $x$-direction (dashed yellow line).}
\end{figure}

Figure~\ref{fig:ionizationVolume}a shows the 2D-projection of the obtained electron/ion density distribution for a pulse of $I_0 = 1.9 \times 10^{13}$~\hl{$\mathrm{W}\,\text{cm}^{-2}$} together with the cylindrical volume used for the plasma simulations. Fig.~\ref{fig:ionizationVolume}b depicts the ionization probability in $x$-direction at $(y,z) = (0,0)$ (solid blue line), in $z$-direction at $(x,y) = (0,0)$ (red dotted line) as well as the normalized atomic density distribution in $x$-direction at $(y,z) = (0,0)$ (dashed yellow line) as a Thomas-Fermi profile of the BEC for the experimental trap frequencies. The atomic cloud is almost spherical as the trapping frequencies are similar in all three dimensions. Whereas the photoionization can be regarded as local in in $x$- and $y$-direction, we fully ionize the atomic ensemble in the direction of the pulse propagation $z$.

\section*{2. Electric Field Configuration}

\begin{figure}[tbh]
 \includegraphics[width=0.8\linewidth]{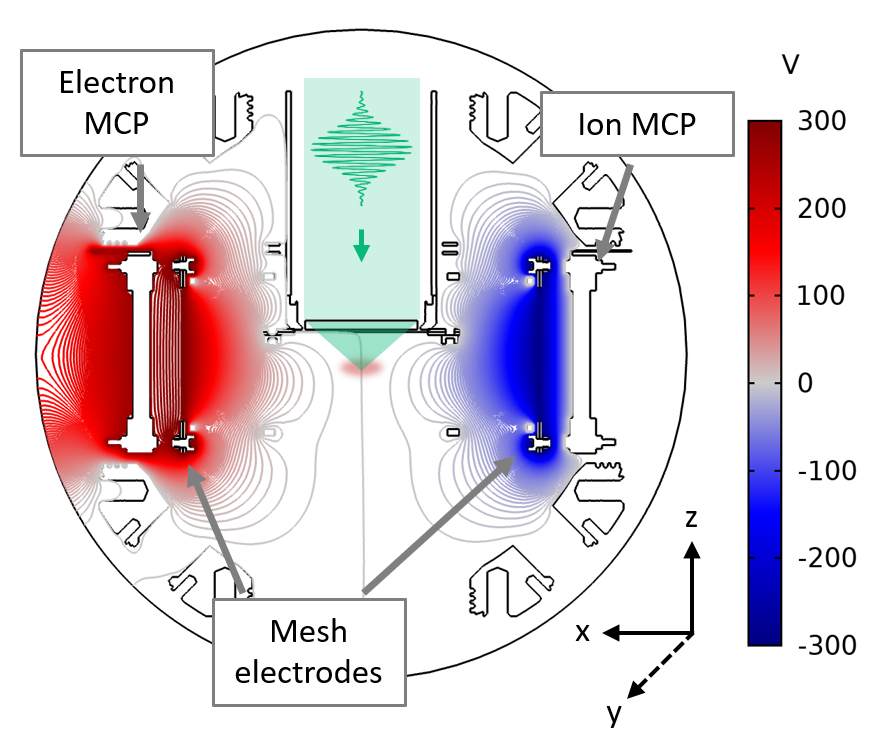}%
 \caption{\label{fig:FieldConfig} \textbf{Electric extraction field.} Sectional view into the 3D-CAD model of the vacuum chamber (black lines). The electric potential obtained for $\pm U_\mathrm{ext} = 300$~V is depicted as equipotential lines (from blue to red). As the ion MCP and the vacuum chamber are grounded, the electric potential is dominated by the mesh electrodes and the electron MCP.}
\end{figure}
We use the Electrostatics Module within the COMSOL Multiphysics\textsuperscript{\textregistered} software \cite{COMSOL} to calculate the electrostatic field configuration for each extraction voltage. For this purpose, we include the 3D computer-aided design (CAD) geometry of our setup into the simulation. The use of finite element methods (FEM) allows for the calculation of the electric potential landscape produced by the different electrodes. We used the physics-controlled mesh option 'finer' with a minimum/maximum element size of 0.88/12~mm.

Figure~\ref{fig:FieldConfig} shows a sectional view into the vacuum chamber together with the equipotential lines obtained for $\pm U_\mathrm{ext} = 300$~V. Since the high-resolution objective needs to be close to the ionization volume, the reentrant window has to be shielded by a grounded high-transparency copper mesh (identical to the ones used for the electrodes) to avoid accumulation of static charge. This leads to a non-linear extraction field strongly increasing towards the detector, as well as a top-down asymmetry, which explains the non-spherical distributions observed experimentally (compare Fig.~3).

%\begin{table}[h!]
%\label{tab:ultracoldplasma_Efield}
%\centering
%\begin{tabular}{l|llll}
%$U_\mathrm{ext}$~(V)  & 5 & 25 & 100 & 300\\
% \hline
%$E_\mathrm{IV}$~(V/m) & 4.6 & 14.4 & 55 & 162\\
%\end{tabular}
%\caption{\textbf{Extraction field strength.} Magnitude of the electric fields in the ionization volume for different extraction voltages.}
%\end{table}

The extraction field in the interaction region can be precisely controlled by the voltages $U_\mathrm{ext}$. However, its amplitude is not perfectly proportional to the applied voltage for low extraction fields ($\pm U_\mathrm{ext} = 5~$V). This is due to the voltages on the electron MCP, which give rise to an electric field on the order of 2 \hl{$\mathrm{V}\,\mathrm{m}^{-1}$} in the center. In addition, the shielding by the vacuum chamber is included in the simulations.

\section*{3. Stray Field Control}

The FEM simulations of the electromagnetic fields provide valuable insight into the level of control over the electric and magnetic stray fields. At an extraction voltage of $\pm U_\mathrm{ext} = 5$~V, which corresponds to an electric field of 4.6 \hl{$\mathrm{V}\,\mathrm{m}^{-1}$} at the center, the extraction field still clearly dominates over the electrical stray fields.
For smaller extraction fields however, our measurements deviate from the theoretical predictions due to electric fields. In our experimental setup, these electric stray fields are passively shielded by the grounded vacuum chamber and electric gradients are controlled down to the \hl{$\mathrm{V}\,\mathrm{m}^{-1}$} level.

Helmholtz coils are used to compensate magnetic fields in all three spatial dimensions with an accuracy of 10 mG. A homogeneous compensation over the extent of the detection units is enabled by meter-sized coils. We have been working here with a magnetic field offset of 370 mG along the $y$-axis, which increases our energy resolution and centers the electron signal onto the detectors.

\section*{4. CPT Plasma Simulations}

\begin{figure}[tbh]
 \includegraphics[width=1\linewidth]{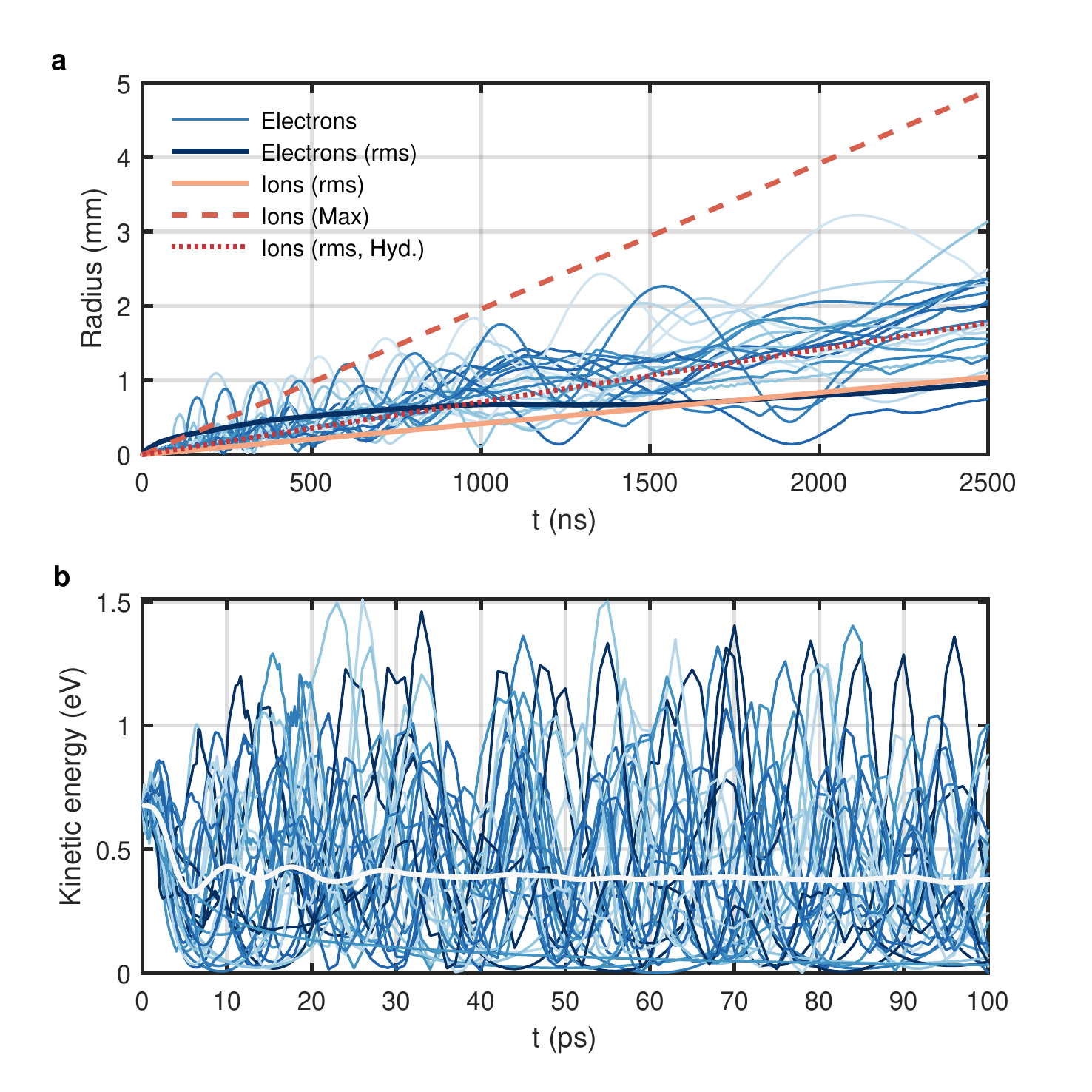}%
 \caption{\label{fig:Trajectories} \textbf{Plasma dynamics at the single particle level.} The dynamics is shown for a generic selection of 31 plasma electrons out of $N_\mathrm{e/i} = 4000$ simulated particles. \textbf{a}. Time-evolution of the distance to the ionization center for the individual plasma electrons (blue lines) in the field-free plasma simulation (see Fig.~4f-g) together with the rms radius of the electron/ion ensemble (bold solid blue line / solid light red line) as well as the maximal ion radius (dashed red line). In addition, the rms ion radius evolution expected from hydrodynamic expansion is given (dark red dotted line). \textbf{b}. Time-evolution of the single electron kinetic energies (blue lines). The mean kinetic energy of all plasma electrons is given by the bold white line.}
\end{figure}

The plasma simulations are based on CPT simulations and include Coulomb interaction between the charged particles (see Methods). Besides quantities such as the mean kinetic energies of the electron/ion ensembles, the CPT plasma simulations provide detailed access to the dynamics of each charged particle. 
Figure~\ref{fig:Trajectories}a shows the time-evolution of the distance from the ionization center of single plasma electrons (blue lines) for the simulation depicted in Fig.~4f-g without extraction field. For clarity, the graph only depicts a random selection of 31 plasma electrons. One clearly identifies an oscillatory motion for the different particles exhibiting different frequencies ranging from hundreds of gigahertz to a few megahertz. The frequency is decreasing with increasing amplitude of the oscillations due to screening by the more closely bound electrons. The maximal ion radius (dashed red line) given by the maximum of all ion distances from the ionization center is used to distinguish between plasma and escaping electrons. In this simulation electrons are regarded as plasma electrons, if their distance at t~=~2500~ns is less than twice the maximum ion radius. 

In addition, the rms electron/ion radius is given for each time step (bold solid blue line / solid light red line). The asymptotic expansion velocity of the rms ion radius of $v_\mathrm{i,rms} = 418~\mathrm{m}\,\mathrm{s}^{-1}$ is in reasonable agreement with the expected plasma expansion velocity $v_\mathrm{hyd} = \sqrt{k_\mathrm{B} \left( T_\mathrm{e,0} + T_\mathrm{i,0} \right) / m_\mathrm{i}} \approx 710~\mathrm{m}\,\mathrm{s}^{-1}$ for hydrodynamic expansion (dark red dotted line) at an initial electron temperature of $T_\mathrm{e,0} = 5250$~K and negligible initial ion temperature $T_\mathrm{i,0}$~\cite{Killian2007a}.

Figure~\ref{fig:Trajectories}b depicts the corresponding kinetic energies of the plasma electrons for the first 100~ps (blue lines) as well as the mean kinetic energy of all 1961 plasma electrons (bold white line). The observed oscillations in the mean kinetic energy are caused by the oscillatory motion of the electrons with similar frequencies and the apparent damping is evoked by their dephasing.

For the simulations including an extraction field, the differentiation between plasma and escaping electrons is more challenging, since electrons enter large orbits, where the non-linear extraction field become dominant, and escape from the plasma. Thus, the extraction fields reduce the number of plasma electrons by about 80\% over time. For the simulations shown in Fig.~4g at $\pm U_\mathrm{ext} = 300$~V and $\pm U_\mathrm{ext} = 5$~V electrons are regarded as plasma electrons, if their distance from the ionization center after 40~ns / 300~ns is less than 1~mm.

\section*{5. Effective Space Charge Potential}

The CPT plasma simulations furthermore provide access to the space charge potential well created by the unpaired ions during the plasma expansion. The extraction field $E_\mathrm{ext}$ adds up to the 1D space charge potential along the detection axis lowering the effective trapping potential

\begin{equation}
 U_\mathrm{eff}(r) = U(r) - E_\mathrm{ext} \cdot r,
 \label{eq:Ueff}
\end{equation}

\noindent where $r$ denotes the distance to the ionization center in the direction of the extraction field. Figure~4g (red lines) shows the evolution of the effective space charge potential depth for the plasma simulations without extraction field as well as for extraction voltages of $\pm U_\mathrm{ext} = 5$~V and $\pm U_\mathrm{ext} = 300$~V (corresponding to $E_\mathrm{ext}$ = 4.6~\hl{$\mathrm{V}\,\mathrm{m}^{-1}$} and $E_\mathrm{ext}$ = 162~\hl{$\mathrm{V}\,\mathrm{m}^{-1}$}). Here, for each time-step $U(r)$ is approximated by the Coulomb potential of a homogeneously charged sphere

\begin{equation}
 U(r) =
 \begin{cases}
  \frac{Q}{8 \pi \epsilon_0 R} \left(3 - \frac{{r}^2}{R^2}\right), & r \le R\\
  \frac{Q}{4 \pi \epsilon_0 r}, & r > R 
 \end{cases}  
 \label{eq:USphere}
\end{equation}

\noindent where the radius $R$ is given by the maximal ion radius and the charge $Q = e \cdot N_{\mathrm{diff}}$ is given by the difference $N_\mathrm{diff}$ of the number of ions and electrons within the maximum ion radius. The potential depth is determined by the difference of the local maximum $U_\mathrm{eff}(r_\mathrm{max})$ and the local minimum $U_\mathrm{eff}(r_\mathrm{min})$ of the effective potential at

\begin{equation}
 r_\mathrm{min} = \frac{4 \pi \epsilon_0 E_\mathrm{ext} R^3}{\text{$Q$}} \text{  and  } r_\mathrm{max} = \sqrt{\frac{Q}{4 \pi \epsilon_0 E_\mathrm{ext}}}.
 \label{eq:rminmax}
\end{equation}

\section*{6. Plasma Formation} 

For plasma formation, the depth of the space charge potential has to exceed the electronic excess energy $E_\mathrm{kin,e}$. Thus, for a given excess energy and a Gaussian spatial distribution of charge carriers, the creation of a minimum ion number $N^* = E_\mathrm{kin,e}/U_0$ is required, where $U_0 = \sqrt{\frac{2}{\pi}} \frac{e^2}{4 \pi \epsilon_0 \sigma}$ and $\sigma$ denotes the rms radius of the ionic distribution~\cite{Killian1999}. The experimentally realized ionic distribution is approximated by a Gaussian distribution with the arithmetic mean radius of $\sigma = (2\times 1.35~\mu\mathrm{m} + 5~\mu\mathrm{m})/3$ leading to a critical ion number of $N^* =960$.

\begin{figure}[t!]
\centering
 \includegraphics[width=1\linewidth]{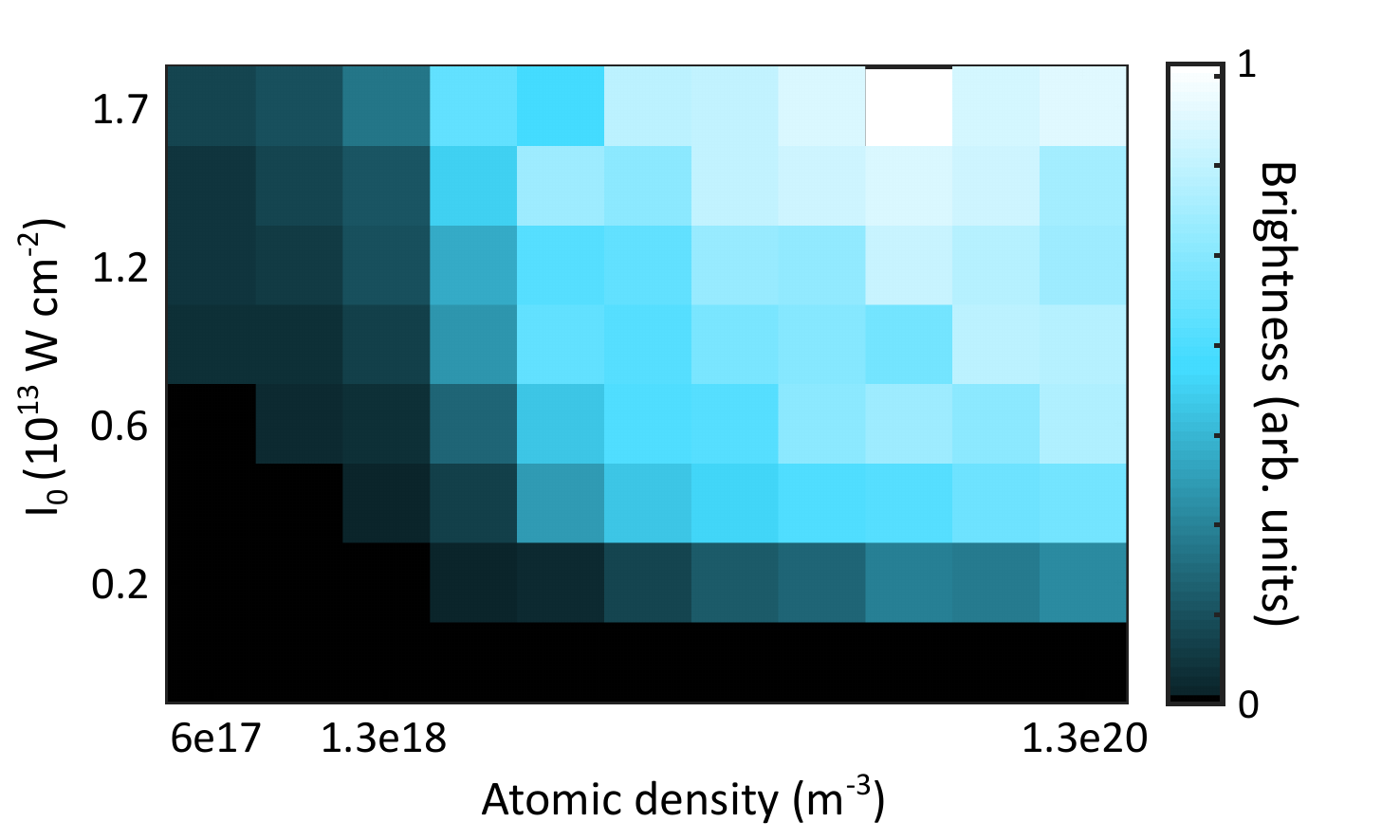}%
 \caption{\label{fig:IntensityDensityScan} \textbf{Critical charge carrier density for plasma formation.} Measured brightness of the electron signal in an elliptical area enclosing low kinetic energy electrons. The atomic target density has been varied over three orders of magnitude and the pulse intensity from $0.1 \times 10^{13}$~\hl{$\mathrm{W}\,\text{cm}^{-2}$} to $1.7 \times 10^{13}$~\hl{$\mathrm{W}\,\text{cm}^{-2}$}.}
\end{figure}

Figure~\ref{fig:IntensityDensityScan} shows the measured brightness at $\pm U_\mathrm{ext} = 300$~V in an elliptical area around the plasma electrons on the detector. The number of plasma electrons, as signature of the plasma formation, displays a strong dependency on the critical charge carrier density. This density has been varied either by the pulse intensity (from $0.1 \times 10^{13}$~\hl{$\mathrm{W}\,\text{cm}^{-2}$} to $1.7 \times 10^{13}$~\hl{$\mathrm{W}\,\text{cm}^{-2}$}) or the atomic density (from $6 \times 10^{17}$ m$^{-3}$ to $\rho$ = $1.3 \times 10^{20}$ m$^{-3}$).

In order to vary the atomic density, we modify the evaporation efficiency, which leads to a reduced number of atoms in the final optical dipole trap. The density is scanned from an ultracold thermal cloud with $\rho$ = $6 \times 10^{17}$ m$^{-3}$ to an almost pure condensate with $\rho$ = $1.3 \times 10^{20}$ m$^{-3}$. The densities are determined by the analysis of the optical density distributions in a time-of-flight measurement recorded by absorption imaging for different timesteps after switching off the optical dipole trap. 

Whereas the calculation of in-situ atomic densities is reliable for the limiting cases of a fully condensed or thermal atomic sample, it is known to be difficult for partly condensed samples as the determination requires a bimodal fit using a Gaussian as well as the Thomas-Fermi density model. However, the critical number of ionized atoms can be extracted from the numbers of detected electrons for each intensity density combination. As the ionization volume slightly increases with increasing peak intensity, the critical number increases as well. While for $I_\mathrm{0} = 1.7 \times 10^{12} $~\hl{$\mathrm{W}\,\text{cm}^{-2}$} around 500 ions need to be created, for $I_\mathrm{0} = 1.9 \times 10^{13} $~\hl{$\mathrm{W}\,\text{cm}^{-2}$} approximately 1000 electrons are required, which agrees well with the expected value of $N^*$ = 960.

\begin{figure*}[tbh]
 \includegraphics[width=1\linewidth]{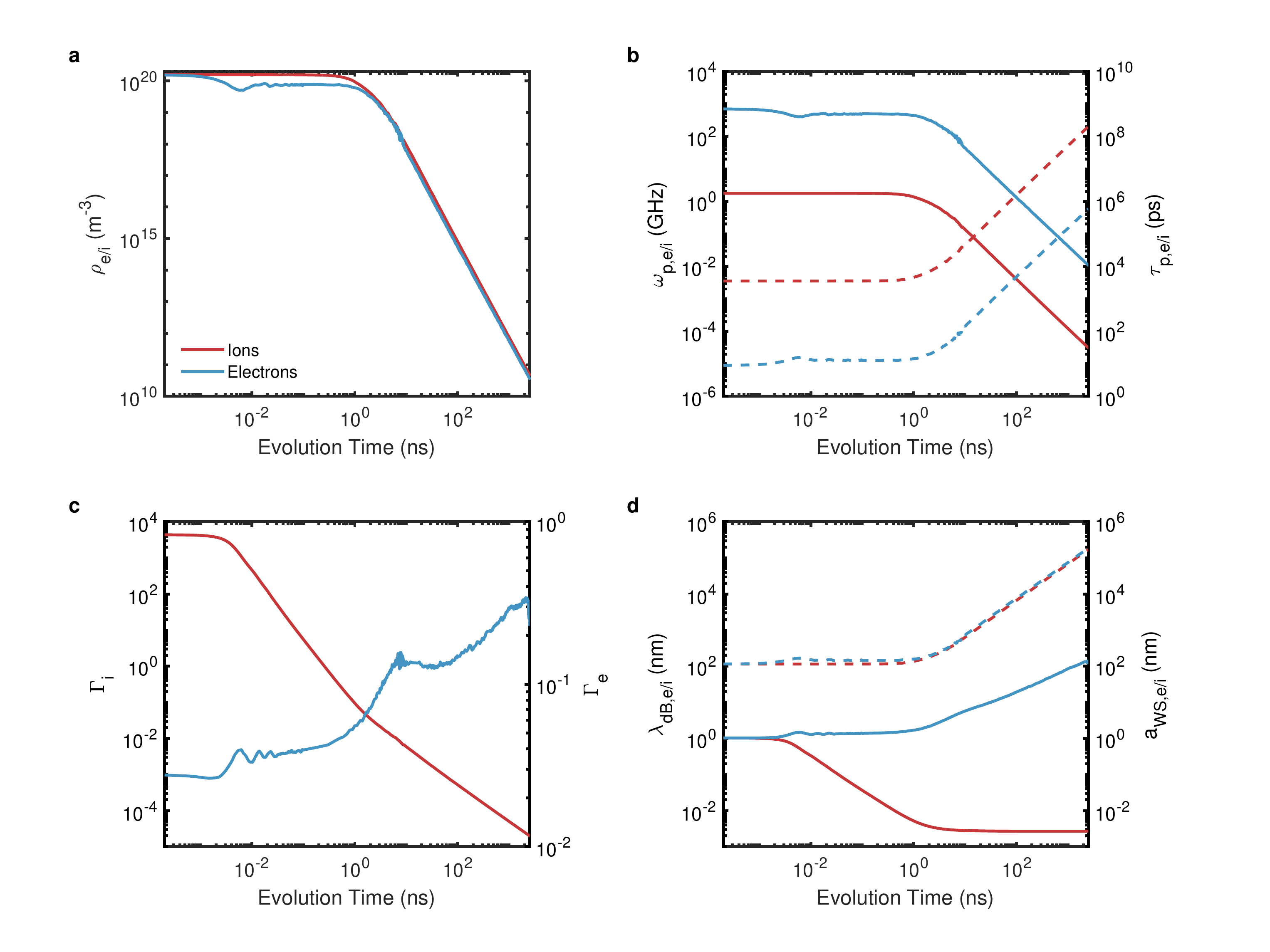}%
 \caption{\label{fig:PlasmaParameters} \textbf{Time-evolution of the simulated plasma parameters.} \textbf{a.} Density of electrons (blue line) and ions (red line) in a spherical volume around the ionization center. \textbf{b.} Electron/ion plasma period $\tau_\mathrm{p,e/i} = 2 \pi \omega_\mathrm{p,e/i}^{-1} $ (dashed blue/red line) and frequencies $\omega_\mathrm{p,e/i}$ (solid blue/red lines). \textbf{c.} Coulomb coupling parameter $\Gamma_\mathrm{e,i}$ of the electron/ion component (blue/red line) determined by the particle densities and the plasma electron/ion mean kinetic energy. \textbf{d.} Electron/ion de Broglie wavelength $\lambda_\mathrm{dB,e/i}$ (solid blue/red line) and Wigner-Seitz radius $a_\mathrm{WS,e/i}$ (dashed blue/red line).}
\end{figure*}

\section*{7. Plasma Parameters}
During the expansion, the densities and kinetic energies of the plasma vary over several orders of magnitude. Figure~\ref{fig:PlasmaParameters} displays the time evolution of central parameters extracted from the CPT plasma simulation. Figure~\ref{fig:PlasmaParameters}a shows the evolution of the electron and ion density in the center of the plasma. The electron density drops within the initial expansion and stays constant for the first nanosecond. As the ionic component expands, the electron and ion density both decrease and even fall below typical initial UNP densities. The density evolution determines the plasma frequency as well as the plasma period given in Fig.~\ref{fig:PlasmaParameters}b. The decrease of the electron density explains the deceleration of the electronic orbital oscillations during the plasma expansion. In addition, the ionic plasma frequency significantly decreases within the first plasma period (given by the initial ionic density). In contrast to UNP, where plasma expansion can be regarded as slow in relation to the inverse ionic plasma frequency, here, the ionic plasma period exceeds the plasma expansion duration, thus preventing ionic thermalization.

In Fig.~\ref{fig:PlasmaParameters}c the ion and electron coupling parameters are depicted. The ionic coupling parameter decreases after the first electron plasma period when charge imbalance is established due to ionic acceleration and increasing interparticle distance. On the contrary, the electronic coupling parameter increases during the plasma expansion since the electron temperature decreases over orders of magnitude. The simulations reveal a maximum coupling parameter of $\Gamma_\mathrm{e} = 0.3$ approaching significant electron coupling.

Electron temperatures in the Kelvin domain raise the question of quantum degeneracy for the electronic ensemble. Fig.~\ref{fig:PlasmaParameters}d illustrates the electron/ion de Broglie wavelengths $\lambda_\mathrm{dB,e/i}$ at different expansion times. Whereas the ionic wavelength quickly decreases, the electrons reach a maximum de Broglie wavelength on the order of $\lambda_\mathrm{dB,e} \approx 100~$nm at the end of the plasma expansion. However, the ratio of mean interparticle distance given by the Wigner-Seitz radius $a_\mathrm{WS,e/i}$ (dashed blue/red line) and the de Broglie wavelength never exceeds 1.3 \%, which yields $E_\mathrm{kin,e}/E_\mathrm{F} > 6000$, where $E_\mathrm{F} = \frac{\hbar^2}{2 m_\mathrm{e}} \left(3 \pi^2 \rho_\mathrm{e}\right)^{2/3}$ denotes the electron Fermi energy with the reduced Planck constant $\hbar$. Thus, a quantum mechanical description required for a fermionic ensemble close to degeneracy can be safely disregarded.

\section*{8. Three-Body Recombination}
\begin{figure}[tbh]
 \includegraphics[width=1\linewidth]{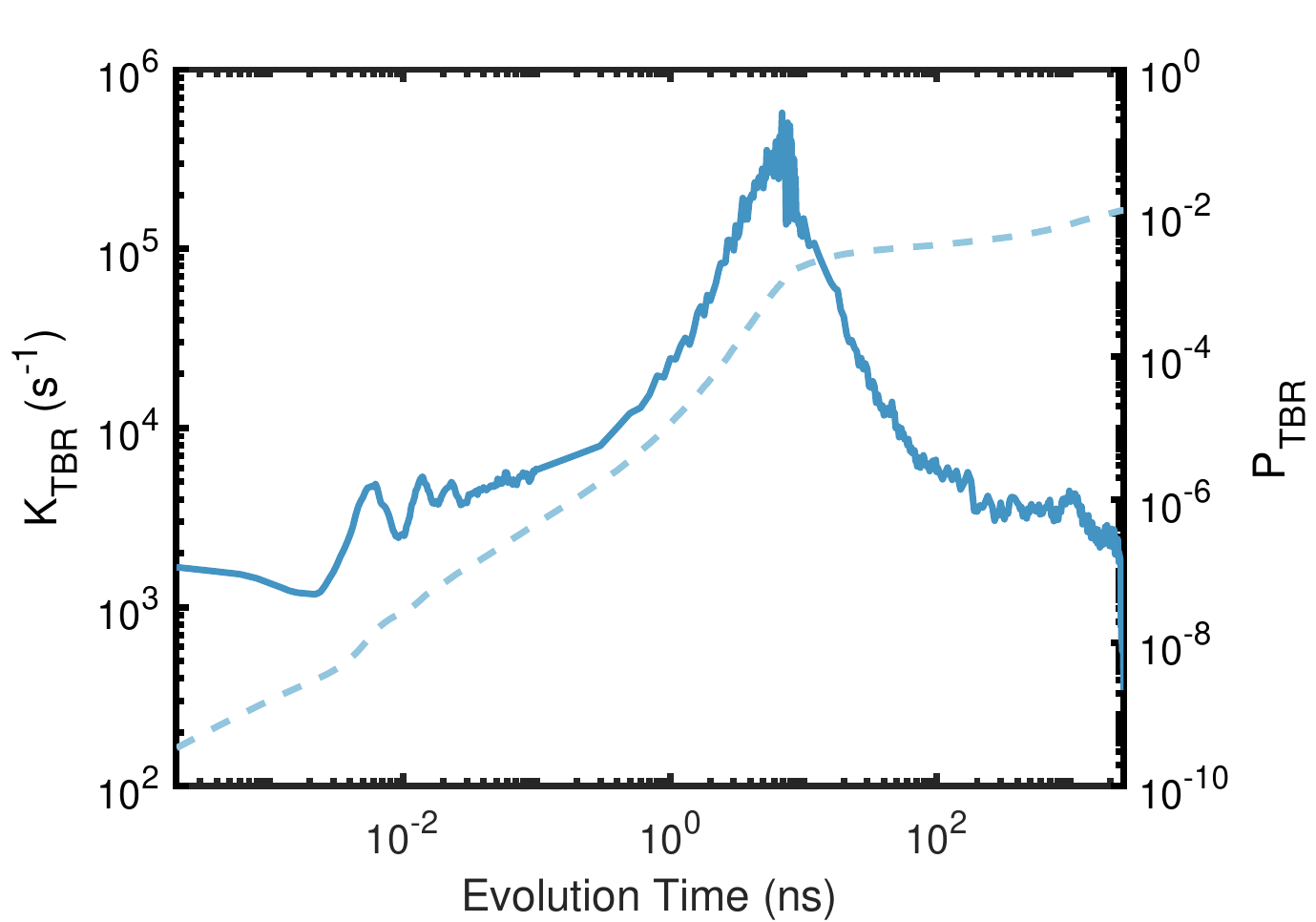}%
 \caption{\label{fig:TBRrate} \textbf{Three-body recombination in the ultracold microplasma.} Time-evolution of the TBR rate $K_\mathrm{TBR}$ per ion during the plasma expansion (solid blue line) as well as the resulting time-integrated recombination probability $P_\mathrm{TBR}$ per ion (dashed blue line).}
\end{figure}

For ultracold plasma in the density and temperature regime described in this manuscript, three-body recombination (TBR) is expected to be the dominant process of electron-ion recombination. The TBR rate $K_\mathrm{TBR}$ per ion according to classical TBR theory is given by $K_\mathrm{TBR} \approx 3.8 \times 10^{-9}~T_\mathrm{e}^{-9/2} \rho_\mathrm{e}$~s$^{-1}$, where $T_\mathrm{e}$ is the electron temperature in K and the electron density $\rho_\mathrm{e}$ is given in cm$^{-3}$~\cite{Killian2007a}. Figure~\ref{fig:TBRrate} shows the calculated TBR rate per ion (solid line) as well as the time-integrated TBR probability per ion (dashed line). After 2.5 µs of plasma expansion, a cumulated TBR probability of approximately 1\% is reached. As a result, the plasma lifetime is expected to be on the order of 100 \textmu s before a significant fraction of Rydberg excitations are created. However, on the ten microsecond timescale, when the plasma is dilute and collisions barely occur, radiative and dielectronic recombination might further limit the plasma lifetime.

\section*{9. Time-resolved Electron Detection}

\begin{figure}[h!]
\centering
 \includegraphics[width=0.90\linewidth]{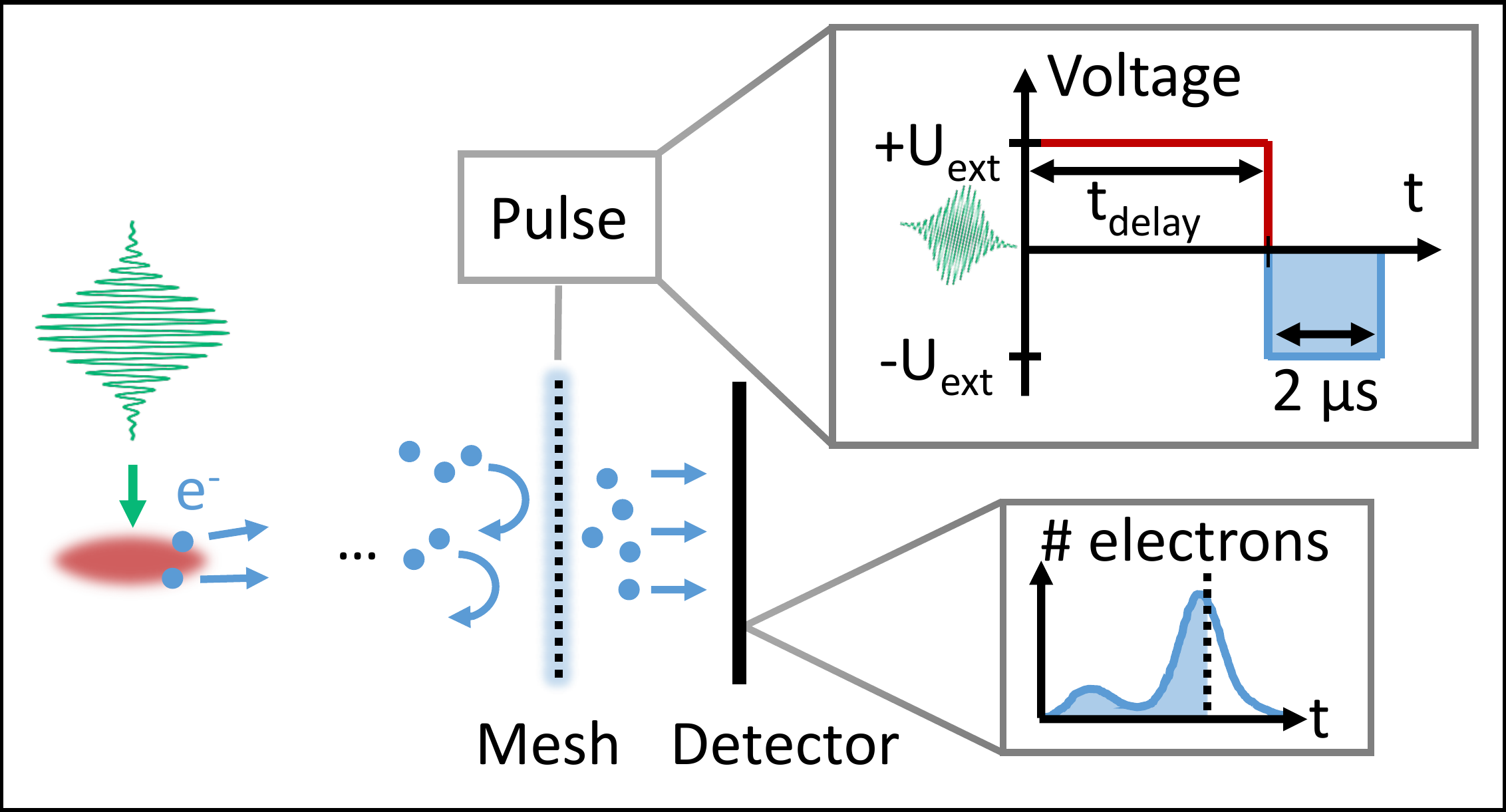}%
 \caption{\label{fig:pulseScheme} \textbf{Photoelectron time-of-flight measurement.} The photoelectrons created in the BEC are accelerated towards the detector by a variable extraction field. After a variable time $t_\mathrm{delay}$, a microsecond voltage pulse switches the polarity of the extraction field. The resulting repulsive electric field prevents electrons with arrival times $t_\mathrm{ToF} > t_\mathrm{delay}$ from passing the mesh. Thus, the detector records the time-integrated electron signal up to $t_\mathrm{delay}$.}
\end{figure}

The experimental setup gives access to the distribution of arrival times of the detected electrons by a gated detection scheme (Fig.~\ref{fig:pulseScheme}). For this purpose, a repulsive voltage pulse is applied to the electron extraction mesh after a variable delay $t_\mathrm{delay}$ after the femtosecond laser pulse. The rapidly switched potential prevents electrons from passing the extraction mesh for time-of-flight durations $\tau_\mathrm{ToF} > t_\mathrm{delay}$. The applied pulse has a duration of 2~\textmu s and is capacitively coupled onto the meshes. This enables to measure the accumulated electron signal up to $t_\mathrm{delay}$, while maintaining the spatial resolution. The estimated timing uncertainty of 30~ns is caused by the temporal jitter and the pulse rise time.

\begin{figure}[h!]
 \includegraphics[width=0.95\linewidth]{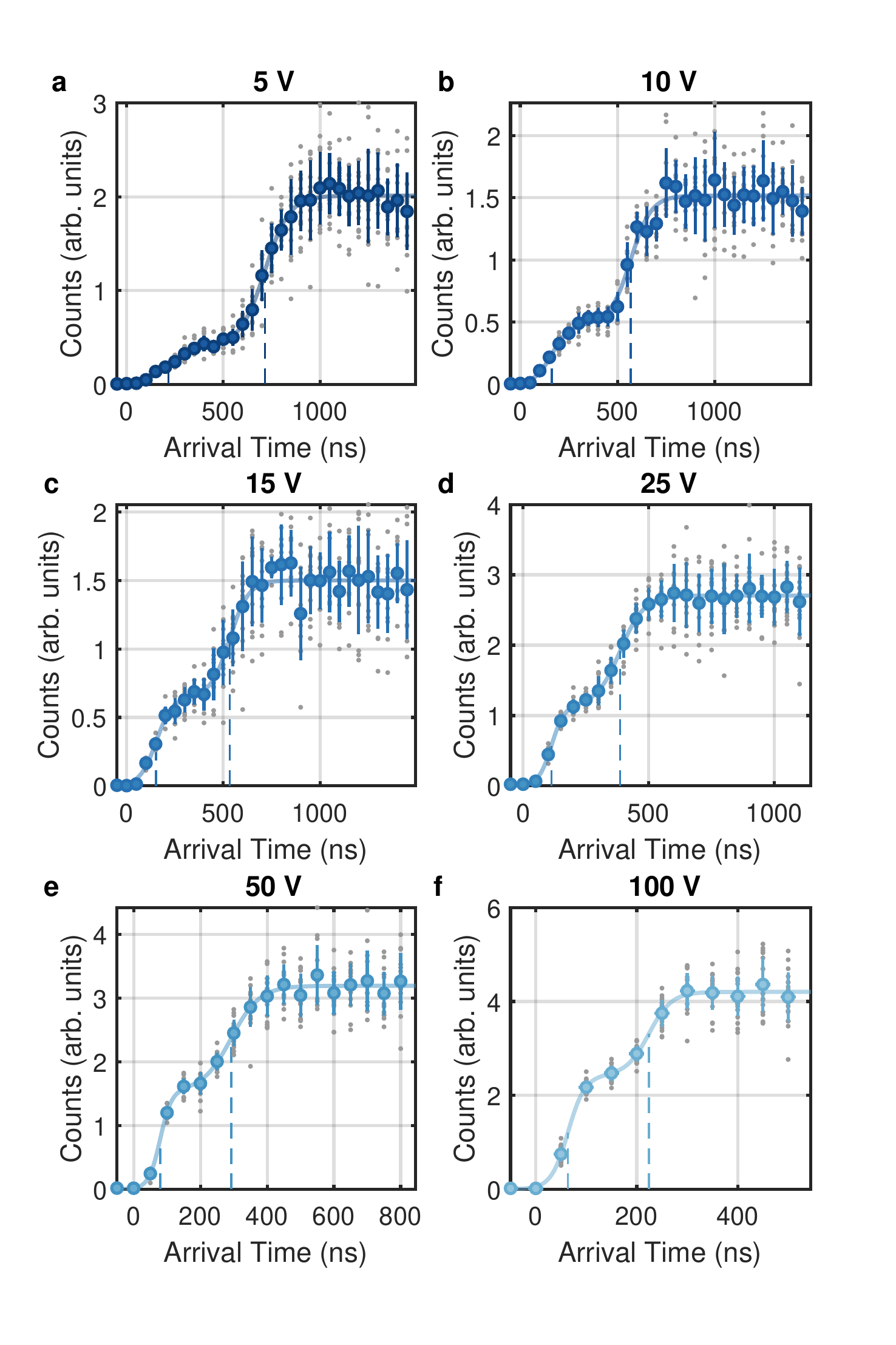}%
 \caption{\label{fig:SigmoidFit} \textbf{Electron arrival time.} \hl{\textbf{a.-f.}} Accumulated electron counts measured at a peak intensity of $1.2 \times 10^{13}$~\hl{$\mathrm{W}\,\text{cm}^{-2}$} for different extraction voltages (see Fig.~5b) \hl{(black data points)}. The vertical error bars \hl{of the binned data (blue data points)} are given by the standard deviation over all realizations and the horizontal ones indicate the time uncertainty of the repulsive voltage pulse. The solid lines show the double-sigmoid functions according to Eq.~(\ref{eq:double_sigmoid}) obtained by a fit to the data points. The dashed lines depict the inflection points $c_1$ and $c_2$ of each sigmoid function.}
\end{figure}

In order to analyze the obtained temporal profiles (see Fig.~5b) quantitatively, a double-sigmoid function

\begin{equation}
 f(t) = \frac{a_1}{1+e^{-b_1\cdot(t-c_1)}} + \frac{a_2}{1+e^{-b_2\cdot(t-c_2)}}
 \label{eq:double_sigmoid}
\end{equation}

\noindent is fitted to the measured data. Figure~\ref{fig:SigmoidFit} shows the fit functions for $\pm U_\mathrm{ext} = 100$~V, $\pm U_\mathrm{ext} = 50$~V, $\pm U_\mathrm{ext} = 25$~V, $\pm U_\mathrm{ext} = 15$~V, $\pm U_\mathrm{ext} = 10$~V and $\pm U_\mathrm{ext} = 5$~V (solid lines, from light blue to dark blue). The two inflection points for the double-sigmoid functions are given by $c_1$ and $c_2$ (vertical dashed lines). The spectra in Fig.~5c are the time derivatives of the fitted functions. The arrival time difference $c_2 - c_1$ determines the plasma lifetime shown in Fig.~5d. The error bars are given by the 95\% confidence interval for the inflection points.

\section*{10. Intensity Calibration}

\begin{figure}[t!]
\centering
 \includegraphics[width=1\linewidth]{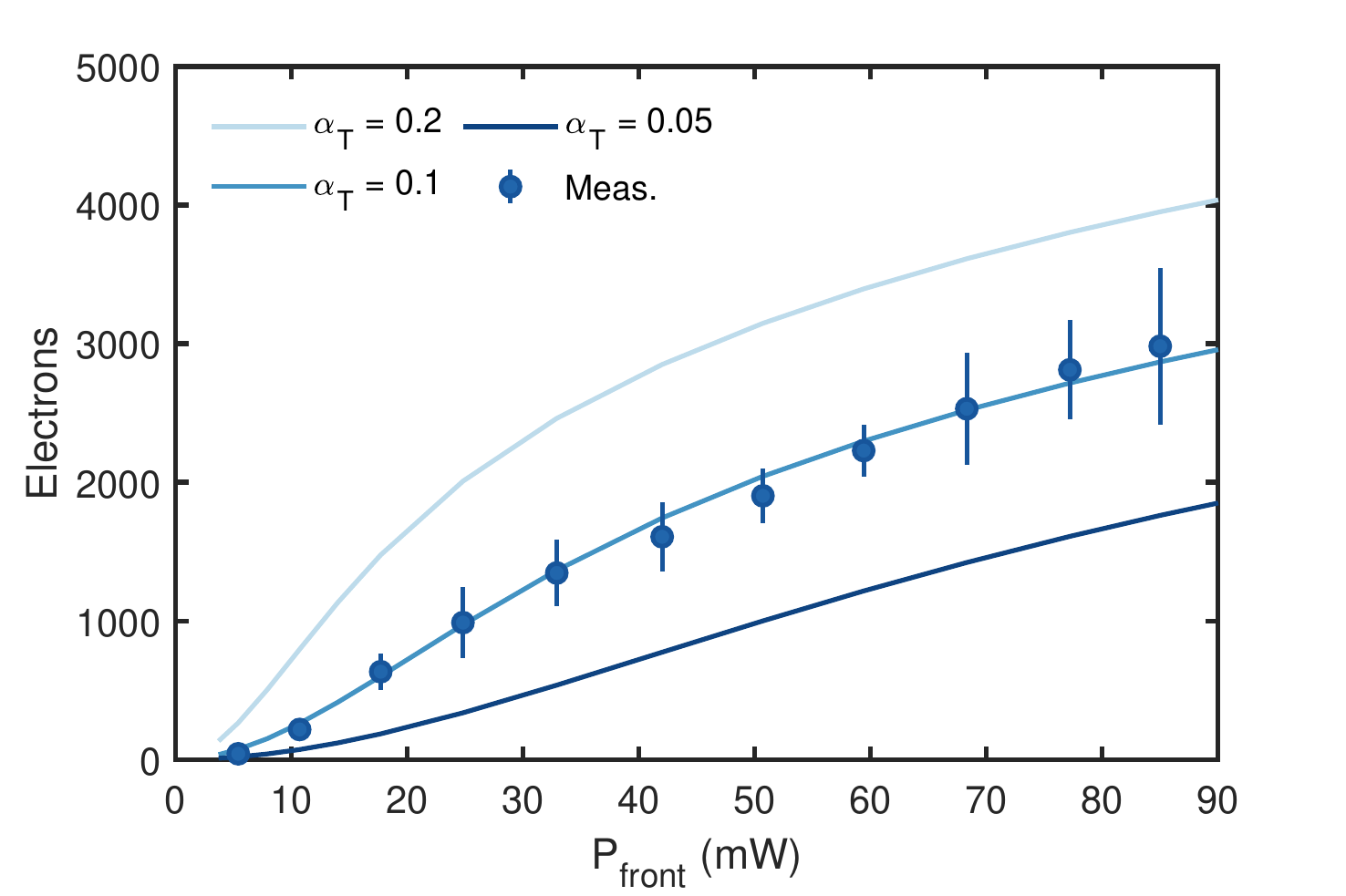}%
 \caption{\label{fig:calibrateIntensity} \textbf{Intensity Calibration}. Measured numbers of electrons for different laser powers $P_\mathrm{front}$ at $\pm U_\mathrm{ext} = 200$~V (data points) as well as the calculated numbers of electrons assuming a transmittance of 0.2, 0.1 and 0.05 (solid lines). \hl{The vertical error bars are given by the standard deviation over all realizations.}}
\end{figure}

The experimental setup only provides access to the femtosecond laser power before passing the high resolution microscope objective. As the transmittance $\alpha_\mathrm{T}$ critically depends on pointing and angle of the incident laser beam, the actual peak intensities inside the vacuum chamber have to be calibrated. The averaged laser power used for the calculation of the applied peak intensity (see Methods) is given by $P = \alpha_\mathrm{T}P_\mathrm{front}$. Here, $P_\mathrm{front}$ denotes the power in front of the objective, which is measured through a circular aperture with the same diameter as the objective aperture (4 mm) at a pulse repetition rate of 100 kHz. Figure~\ref{fig:calibrateIntensity} shows the measured number of electrons for different powers $P_\mathrm{front}$ at an extraction voltage of $\pm U_\mathrm{ext} = 200$~V (data points). The solid lines depict the expected numbers of electrons assuming a transmittance of 0.2, 0.1 and 0.05, which are calculated by use of the absolute ionization probabilities reported in~\cite{Wessels2018} as well as the beam waist measured with an identical objective (see Methods). The best agreement is obtained for $\alpha_\mathrm{T} = 0.1$.

\section*{11. Ultracold Electron Source}

The electron cooling mechanisms in ultracold plasma can be exploited for plasma-based ultracold electron sources~\cite{Claessens2005} producing low-emittance electron bunches. These bunches can be used to seed high-brilliance particle accelerators~\cite{McCulloch2011} and for coherent imaging of biological systems~\cite{Speirs2015}.
With the final electron temperature of $T_\mathrm{e}$~$\approx$~10~K and an rms electron bunch radius $\sigma_\mathrm{r}$= 0.52 mm, we achieve a normalized rms emittance of $\epsilon_\mathrm{r} = \sigma_\mathrm{r}\cdot\sqrt{k_\mathrm{B} T_\mathrm{e} / m_\mathrm{e} c^2} = 21$ nm rad and a relative transverse coherence length $C_{\perp} = \hbar/(m_\mathrm{e} c \epsilon_\mathrm{r}) = 2 \times 10^{-5}$. In our experimental setup, the electron excess energy can be reduced further by working closer to the ionization threshold, allowing to approach the value of $C_{\perp} = 10^{-3}$ required for single-shot electron diffraction~\cite{Engelen2013}.

\end{document}